\documentclass[fleqn,usenatbib]{mnras}

\usepackage{booktabs,caption}
\usepackage{arydshln}
\usepackage[flushleft]{threeparttable}

\usepackage[T1]{fontenc}

\DeclareRobustCommand{\VAN}[3]{#2}
\let\VANthebibliography\thebibliography
\def\thebibliography{\DeclareRobustCommand{\VAN}[3]{##3}\VANthebibliography}

\usepackage{graphicx}	% Including figure files
\usepackage{amsmath}	% Advanced maths commands
\usepackage{amssymb}	% Extra maths symbols
\usepackage{multirow}
\usepackage{newtxtext,newtxmath}
\defcitealias{Buzzo_22b}{B22}
\defcitealias{Ferre-Mateu_23}{FM23}
\title[The stellar populations of the MATLAS UDGs]{Constraining the stellar populations of ultra-diffuse galaxies in the MATLAS survey using spectral energy distribution fitting}

\author[M. L. Buzzo et al.]{Maria Luisa {Buzzo}$^{1,2}$\thanks{E-mail: lgomesbuzzo@swin.edu.au},
Duncan A. Forbes$^{1,2}$,
Thomas H. Jarrett$^{3}$,
Francine R. Marleau$^{4}$,
Pierre-Alain Duc$^{5}$,
\newauthor 
Jean P. Brodie$^{1,2,6}$,
Aaron J. Romanowsky$^{7,8}$,
Jonah S. Gannon$^{1,2}$,
Steven R. Janssens$^{1,2}$,
Joel Pfeffer$^{1,2}$,
\newauthor
Anna {Ferré-Mateu}$^{9,10,1}$,
Lydia Haacke$^{1,2}$,
Warrick J. Couch$^{1}$, 
Sungsoon Lim$^{11}$, and
Rub\'en {S\'anchez-Janssen$^{12}$}
\\ \\
$^{1}$ Centre for Astrophysics and Supercomputing, Swinburne University, John Street, Hawthorn VIC 3122, Australia \\
$^{2}$ ARC Centre of Excellence for All Sky Astrophysics in 3 Dimensions (ASTRO 3D), Australia \\
$^{3}$ Department of Physics and Astronomy, University of the Western Cape, Robert Sobukwe Road, Cape Town, 7535, South Africa \\
$^{4}$ Institut für Astro- und Teilchenphysik, Universität Innsbruck, Technikerstraße 25/8, Innsbruck, A-6020, Austria \\
$^{5}$ Universit\'e de Strasbourg, CNRS, Observatoire astronomique de Strasbourg, UMR 7550, F-67000 Strasbourg, France \\
$^{6}$ University of California Observatories, 1156 High Street, Santa Cruz, CA 95064, USA \\
$^{7}$ Department of Physics and Astronomy, San José State University, One Washington Square, San Jose, CA 95192, USA \\
$^{8}$ Department of Astronomy \& Astrophysics, University of California Santa Cruz, 1156 High Street, Santa Cruz, CA 95064, USA \\
$^{9}$ Instituto Astrofisica de Canarias, Av. Via Lactea s/n, E38205 La Laguna, Spain \\
$^{10}$ Departamento de Astrofisica, Universidad de La Laguna, E-38200, La Laguna, Tenerife, Spain \\
$^{11}$ Department of Astronomy, Yonsei University, 50 Yonsei-ro Seodaemun-gu, Seoul, 03722, Republic of Korea \\ 
$^{12}$ UK Astronomy Technology Centre, Royal Observatory, Blackford Hill, Edinburgh, EH9 3HJ, UK
}

\date{Accepted 2024 February 16. Received 2024 January 19; in original form 2023 October 10}

\pubyear{2024}

\begin{document}
\label{firstpage}
\pagerange{\pageref{firstpage}--\pageref{lastpage}}
\maketitle

\begin{abstract}
We use spectral energy distribution (SED) fitting to place constraints on the stellar populations of 59 ultra-diffuse galaxies (UDGs) in the low-to-moderate density fields of the MATLAS survey. We use the routine \texttt{PROSPECTOR}, coupled with archival data in the optical from DECaLS, and near- and mid-infrared imaging from \textit{WISE}, to recover the stellar masses, ages, metallicities and star formation timescales of the UDGs. 
We find that a subsample of the UDGs lies within the scatter of the mass--metallicity relation (MZR) for local classical dwarfs. However, another subsample is more metal-poor, being consistent with the evolving MZR at high-redshift.
We investigate UDG positioning trends in the mass--metallicity plane as a function of surface brightness, effective radius, axis ratio, local volume density, mass-weighted age, star formation timescale, globular cluster (GC) counts and GC specific frequency. We find that our sample of UDGs can be separated into two main classes. Class A: Comprised of UDGs with lower stellar masses, prolonged star formation histories (SFHs), more elongated, inhabiting less dense environments, hosting fewer GCs, younger, consistent with the classical dwarf MZR, and fainter. Class B: UDGs with higher stellar masses, rapid SFHs, rounder, inhabiting the densest of our probed environments, hosting on average the most numerous GC systems, older, consistent with the high-redshift MZR (i.e., consistent with early-quenching), and brighter. The combination of these properties suggests that UDGs of Class A are consistent with a `puffed-up dwarf' formation scenario, while UDGs of Class B seem to be better explained by `failed galaxy' scenarios. 
\end{abstract}


%%%%%%%%%%%%%%%%%%%%%%%%%%%%%%%%%%%%%%%%%%%%%%%%%%

%%%%%%%%%%%%%%%%% BODY OF PAPER %%%%%%%%%%%%%%%%%%

\section{Introduction}
\label{sec:introduction}

Ultra-diffuse galaxies (UDGs), despite having been heavily studied for almost a decade now \citep{vanDokkum_15}, are still the topic of much debate regarding their formation mechanisms, including dark matter content \citep[e.g.,][]{Toloba_18,ManceraPina_19,ManceraPina_19b,ManceraPina_22,Gannon_20,Gannon_21,Gannon_22,Forbes_20a,Zaritsky_23,Gannon_23,Toloba_23}, stellar populations \citep[e.g.,][]{Roman_Trujillo_17,Ferre-Mateu_18,Ruiz-Lara_18,Chilingarian_19,Barbosa_20,Villaume_22, Buzzo_22b,Heesters_23,Ferre-Mateu_23}, and their globular cluster (GC) systems \citep[e.g.,][]{vanDokkum_18,Trujillo_19,Gannon_21,Gannon_22,Gannon_23,Forbes_20a,Muller_20, Danieli_22}.

While resembling classical dwarf galaxies in terms of their stellar masses ($M_{\star} < 10^9 M_{\odot}$) and surface brightnesses ($\mu_{g,0} > 24$ mag arcsec$^{-2}$), they by definition exhibit much larger effective radii ($R_{\rm e} > 1.5$ kpc). Some of them were found to have populous GC systems and evidence for more massive dark matter halos ($M_{\rm halo} > 10^{11}$ $M_{\odot}$) than their classical dwarf counterparts \citep{Beasley_16,vanDokkum_19b,Forbes_20a,Gannon_20,Gannon_22,Gannon_23,Toloba_23,Zaritsky_23}.
This mixture of dwarf- and massive-like galaxy properties has led to many different proposed formation scenarios for these galaxies. 

A scenario relying on internal processes suggests that UDGs start their lives as classical dwarf galaxies, and through an extended sequence (i.e., repeated episodes over a long period of time) of supernovae feedback \citep{diCintio_17,Chan_18}, they get enlarged to the sizes we observe today. Another scenario suggests that dwarfs with high-spin halos can evolve into UDGs \citep{Amorisco_16,Benavides_21}. We refer to such scenarios as `puffed-up dwarf' formation scenarios.
Other scenarios rely on external processes to form UDGs via e.g., tidal stripping and heating \citep[e.g., ][]{Carleton_19}, tidal interaction \citep[e.g.,][]{Tremmel_20}, which account for the apparent radial size excess. Some UDGs in the past have been suggested to be consistent with a tidal dwarf galaxy (TDG) origin as well \citep[see e.g., ][]{Duc_14, Buzzo_23a}. We refer to this formation scenario throughout the paper as `tidal UDGs'.

In addition to these, one scenario suggests that UDGs started their lifetimes destined to be large and massive but had their star formation truncated early on. As a consequence, their stellar masses do not increase at the same rate as their halos, resulting in galaxies with dwarf-like stellar masses enclosed by halo masses similar to those of more massive M33-like galaxies \citep{vanDokkum_15,Peng_16,Danieli_22, Gannon_23}. For the remainder of the manuscript, we refer to this formation scenario as `failed galaxy' scenario.

One crucial discriminant between these formation scenarios is the stellar populations of UDGs. For example, if they were formed by a puffed-up dwarf scenario, they would overall be expected to have similar stellar populations to those of classical dwarfs. One expectation, for example, is that they may follow the same scaling relations as classical dwarfs, e.g., the mass--metallicity relation (MZR). It is important to mention, however, that different puffed-up dwarf scenarios may lead to very different star formation histories (SFH), i.e., supernovae feedback would lead to a bursty SFH, while high-spin halos or tidal interactions may be more similar to an exponentially decaying SFH, but in both cases the UDGs are expected to preserve the stellar populations of their progenitors. Tidal UDGs are expected to be young, metal-rich, and gas-dominated for their stellar mass \citep{Haslbauer_19, Duc_14}. On the other hand, if UDGs are failed galaxies, the predictions for their stellar populations are less clear. They are not expected to have stellar populations similar to classical dwarfs (as classical dwarfs were not their progenitors), nor to more massive M33-like galaxies (as they did not evolve to become such). The only stellar population prediction for galaxies formed by this scenario is that they would have old or even ancient stellar populations ($\gtrsim$ 10 Gyr), as they by definition have suffered from early-quenching. For the same reason, they may be expected to have lower metallicities and possibly be alpha enhanced as they have had shorter timescales to form stellar masses comparable to those of classical dwarfs that had prolonged star formation histories \citep[see e.g.,][]{Forbes_20a}, and also because, as suggested by \cite{Danieli_22}, these galaxies could be mainly formed of disrupted GCs, thus being more metal-poor. As a key point in the comparison, the MZR is known to evolve with redshift \citep{Ma_15,Chartab_24}. Galaxies that follow the MZR at high-redshift have stellar populations consistent with the chemical enrichment up to that point, thus, either galaxies at high-redshift themselves or galaxies that have not had much chemical evolution since then (i.e., early quenched). We thus may expect failed galaxies to follow the MZRs of high-redshift galaxies.

Some of these predictions were recently confirmed by \cite{Ferre-Mateu_23} (hereafter, \citetalias{Ferre-Mateu_23}), who used the largest spectroscopic study of UDGs to date (25) to show that there is a correlation between $\alpha$-enrichment and the star formation history (SFH) of UDGs, further emphasised by the positioning of UDGs in the cluster phase-space diagram. They found that early-infall UDGs are the ones with the fastest star formation episodes and have higher [$\alpha$/Fe], consistent with early quenching scenarios, such as failed galaxy ones. Conversely, UDGs with prolonged star formation histories are the late infallers, having lower [$\alpha$/Fe] ratios, being better explained by puffed-up dwarf scenarios. 

Additionally, \cite{Buzzo_22b} (hereafter, \citetalias{Buzzo_22b}) used SED fitting to show that there is a correlation between the age, metallicity and environment that UDGs reside in. The older ones display the lowest metallicities, are consistent with the MZR at high-redshift ($z\sim2$) and are found in the densest environments. The younger ones ($\leq$ 8 Gyr) follow the classical dwarf MZR, being more metal-rich and are found in less dense environments, such as the field and groups. 
These findings are in general agreement with the findings of \cite{Barbosa_20}, who used SED fitting to study 100 UDGs in the field and to show that they are mostly young ($\sim 6$ Gyr) and follow the classical dwarf MZR. In their sample, some UDGs were found to be more metal-rich than the classical dwarf MZR ([Fe/H] $\geq -0.5$ dex) and extremely young ($\leq$ 1 Gyr), similar to what has been found for some cluster UDGs \citep{Ferre-Mateu_18,Ruiz-Lara_18}. Their ages and metallicities are consistent with them being tidal UDGs \citep{Collins_Read_22}.

\citetalias{Buzzo_22b} have also shown that the stellar populations of UDGs seem to correlate with their GC richness. Old and metal-poor UDGs have on average the highest number of GCs, while younger and more metal-rich ones (consistent with the classical dwarf MZR) have the lowest GC numbers. This trend, however, was not found by \citetalias{Ferre-Mateu_23} using spectroscopy. Selection effects can be the cause of the differences, since both samples are not fully representative of the population of UDGs. Thus, to understand if this trend holds and whether it can or cannot be extrapolated to other UDGs, it is necessary to test it on a larger sample of UDGs. These would preferably be spread across different environments and span a variety of GC numbers so that selection effects can be diminished and conclusive interpretations made.

In this study, we extend the work of \citetalias{Buzzo_22b} to a sample of 59 UDGs in the MATLAS low- and moderate-density environments \citep{Marleau_21}, to help balancing the sample that was biased towards higher density environments in \citetalias{Buzzo_22b}, and to thus start building up a representative sample of UDGs. 38 out of these 59 MATLAS UDGs have GC counts from \textit{Hubble Space Telescope (HST)} imaging.
Similar to \citetalias{Buzzo_22b}, we employ optical to mid-infrared spectral energy distribution (SED) fitting to explore the stellar populations of these MATLAS UDGs. Buzzo et al. in prep. will combine both samples (i.e., \citetalias{Buzzo_22b} and this study) with a control sample of classical dwarf galaxies to perform an statistically meaningful and comprehensive study of UDGs spread across environments.

The paper is structured as follows: in Section \ref{sec:data} we present a summary of the UDG sample studied in this work and the data available for each UDG. In Section \ref{sec:analysis} we describe our methods to obtain the photometry and morphology of the galaxies and our SED fitting methodology. In Section \ref{sec:results} we provide our results. In Section \ref{sec:discussion} we discuss the implications of our results within the theoretical predictions for UDGs and as compared to the literature. In Section \ref{sec:conclusions} we present the summary and the conclusions of the paper. 

Throughout this paper, when converting distances to redshifts (or vice-versa), we assume the cosmological parameters from the Planck 2020 collaboration \citep[$H_0 = 67.4 \pm 0.5$ km s$^{-1}$ Mpc$^{-1}$; $\Omega_m  = 0.315 \pm 0.007$,][]{Planck_20}.

\section{Data sample}
\label{sec:data}

The MATLAS UDGs were identified by the deep optical imaging of the large observing program Mass Assembly of early-Type gaLAxies with their fine Structures (MATLAS) survey \citep{Duc_14,Duc_15,Duc_20,Bilek_20,Poulain_21,Marleau_21}. 
The sample of 59 UDGs used in this work was selected by \cite{Marleau_21}, using the same UDG criteria as described above (i.e., $\mu_{g,0} > 24$ mag arcsec$^{-2}$, $R_{\rm e} > 1.5$ kpc). These UDGs are all in low-to-moderate density environments. There are no UDGs in high-density environments such as clusters of galaxies in our sample. 
The environment that the galaxies reside in was calculated using a K-nearest neighbour algorithm. Using the 10 nearest galaxies to the UDGs, the local surface density ($\Sigma_{10}$) and local volume density ($\rho_{10}$) are calculated as a proxy for the environment. It therefore does not separate central from satellite UDGs. More details about the environment determination can be found in \cite{Duc_14} and \cite{Marleau_21}. 

For simplicity, throughout this paper we follow the MATLAS convention and refer to the closest massive galaxy to the UDGs as hosts, but that does not necessarily imply that the UDGs are satellites of the massive nearby galaxy. 

No selection against star-forming UDGs was made, meaning that our sample has both star forming and quiescent UDGs, differently from the sample of \citetalias{Buzzo_22b} and \citetalias{Ferre-Mateu_23}, who focused only on quiescent UDGs.

As previously mentioned, the identification of the low surface brightness sources was carried out on the imaging fields around massive early-type galaxies (ETG) in the nearby Universe. The UDGs are, thus, classified as being part of the group/in the vicinity of the massive ETG. However, since the UDG definition is based on physical size, it is imperative to understand the real distance to the galaxy to confirm whether or not it is a bonafide UDG. Few galaxies to date have been followed-up spectroscopically to test if they are indeed associated with the group or suffering from projection effects. The largest follow-up was carried out by \cite{Heesters_23} who studied 56 MATLAS dwarfs, out of which 3 were UDGs (MATLAS-585, MATLAS-2019 and MATLAS-2103). They found that 75\% of the studied galaxies were at the same distances as their hosts, including the three studied UDGs. 
In this study, in addition to the SED fitting, we present spectroscopic follow-up of 3 MATLAS UDG candidates with Keck/DEIMOS (MATLAS-342, MATLAS-368 and MATLAS-1059) to recover their redshifts and probe their association with the group/massive neighbour. The spectroscopic observations and results are described in Appendix \ref{sec:appendix_deimos}. We find that all three UDGs are at the same redshifts as their hosts, in agreement with the findings of \cite{Heesters_23}. We therefore consider it reasonable to assume that the rest of the sample is also at the same redshift as their hosts. We further discuss the assumed redshifts for the MATLAS UDGs in Section \ref{sec:redshifts}.

The description of the imaging data used to perform SED fitting is in Section \ref{sec:phot}. The data used to perform the GC number counts are described in Section \ref{sec:gc_counts}.

\subsection{Photometric data}
\label{sec:phot}
In this work, we use data from the optical to mid-infrared to study the stellar population properties of 59 UDGs in the MATLAS survey. Below we present the data used for each galaxy, along with how the photometry was measured in each band. We note that we tried to recover the photometry in the ultraviolet for the MATLAS UDGs using \textit{Galaxy Evolution Explorer (GALEX)} imaging. However, all of our measurements turned out to be fainter than the 3$\sigma$ limit of the survey, independently if the imaging was observed with the Deep (DIS, $3\sigma_{\rm DIS} (\rm NUV) = 24.9$ mag), Medium (MIS, $3\sigma_{\rm MIS} (\rm NUV) = 23.2$ mag) or All-Sky (AIS, $3\sigma_{\rm AIS} (\rm NUV) = 21.3$ mag) survey \citep{Martin_05,Morrissey_07}.

Optical, near- and mid-IR magnitude measurements are in AB magnitudes and were corrected for Galactic extinction using the two-dimensional dust maps of \citealt{SFD} (recalibrated by \citealt{Schlafly_11}) and the extinction law of \cite{Calzetti_00}.

\subsubsection{DECaLS optical imaging}
\label{sec:DECaLS_data}

Although the MATLAS UDGs were identified using Canada-France-Hawaii Telescope \citep[CFHT, surface brightness limit of 28.5 - 29 mag arcsec$^{-2}$ in the $g$-band,][]{Duc_14, Poulain_21} data, in this study, we use data from the Dark Energy Camera Legacy Survey \citep[DECaLS, surface brightness limit of 28.5 mag arcsec$^{-2}$ in the $r$-band, ][]{Jiaxuan_22} to perform the SED fitting. This is because DECaLS has imaging available in the $g$, $r$, $i$ and $z$ bands for most galaxies in our sample, while CFHT has only the $g$ and $r$ bands (with a few rare cases where the $i$ band is also available). We caution, nonetheless, that the DECaLS coadded imaging is noisy in the outskirts of galaxies \citep{Jiaxuan_22}, which may lead to slight differences in the recovered photometry and morphological parameters when compared to CFHT, although consistent within uncertainties. 

Differently from \citetalias{Buzzo_22b}, in this study we do not use aperture photometry of our galaxies. Interested in obtaining total magnitudes in addition to morphological properties of the galaxies, we use the multi-wavelength galaxy fitting \citep[\texttt{GALFITM,}][]{Haussler_13,Vika_13} routine to study the galaxies in the optical. Detailed explanation of the input parameters and configuration of \texttt{GALFITM} are given in Section \ref{sec:galfitm}.

For all of the galaxies, archival optical coadded data were obtained from the Dark Energy Camera Legacy Survey \citep[DECaLS,][]{Dey_19}. 36 out of our 59 UDGs were observed as part of the DECaLS Data Release 10, which includes the $g$, $r$, $i$ and $z$ bands. The remaining 23 galaxies were observed as part of the DECaLS Data Release 9 and have imaging available in the $g$, $r$ and $z$ bands. The reduction and calibration of the DECaLS data are described in \cite{Dey_19}. 

As discussed in \citetalias{Buzzo_22b}, the coadded DECaLS data have shallower depths and more uncertain sky subtractions than other optical surveys focused on low surface brightness galaxies. Nonetheless, tests carried out by \citetalias{Buzzo_22b} have shown that the photometry obtained with the DECaLS coadded data are consistent within 1 $\sigma$ with the photometry obtained by \cite{Lim_20} and \cite{Pandya_18} using deeper data reduced with an LSB-appropriate pipeline. The photometry recovered from DECaLS is, however, on average 0.1 magnitudes fainter than the ones in the literature. This systematic difference was incorporated into our final magnitudes as described in \citetalias{Buzzo_22b}. 
The final photometry in the optical for all MATLAS UDGs was compared to that obtained by \cite{Poulain_21} using CFHT data and was found to provide similar results.

\subsubsection{\textit{WISE} near-IR and mid-IR imaging}
\label{sec:WISE_data}

The \textit{Wide-field Infrared Survey Explorer} \citep[\textit{WISE}, ][]{Wright_10} is a space telescope that has imaged the entire sky in four filters with effective wavelengths of 3.4, 4.6, 12 and 22 $\mu$m (near to mid-infrared). For this study, we gathered \textit{WISE} data for all the galaxies in our sample. These data are a mix of archival ALLWISE data and bespoke data construction and analysis, including custom mosaic construction from \textit{WISE} single frames. The reduction, calibration and photometric measurement processes are thoroughly described in \citetalias{Buzzo_22b}. We use \textit{WISE} in all of our SED fits as it has been shown by \citetalias{Buzzo_22b} to significantly improve the stellar population results and to help breaking the age--metallicity degeneracy.
All photometric measurements can be found in Table \ref{tab:photometry}.

\subsection{Globular Cluster Numbers}
\label{sec:gc_counts}
Total GC numbers have been obtained for 38 galaxies in our sample using \textit{Hubble Space Telescope (HST)}/ACS imaging. The reduction, source detection, GC candidate selection and final GC counts are thoroughly discussed in Marleau et al. subm. Here we use the total (background and completeness corrected) GC numbers of the MATLAS UDGs to test the trend between GC--richness and metallicity found by \citetalias{Buzzo_22b} on our more complete sample.
In \citetalias{Buzzo_22b} and \citetalias{Ferre-Mateu_23}, UDGs considered to be GC--rich were those with more than 20 GCs, as this value roughly corresponds to a halo mass of $10^{11} M_{\odot}$ \citep{Burkert_Forbes_20}. UDGs with less than 20 GCs were considered GC--poor. In this study, we do not impose this hard separation at $N_{\rm GC} = 20$. Instead, we analyse GC numbers in a continuous manner so that evolutionary trends can be identified.

\section{Analysis}
\label{sec:analysis}

\subsection{\texttt{GALFITM}}
\label{sec:galfitm}
In this study, we use \texttt{GALFITM} \citep{Haussler_13,Vika_13} to perform simultaneous galaxy fitting of our optical bands. This technique allows the combination of lower signal-to-noise (S/N) images, such as in the $z$-band, with images with high S/N, and balances them out, resulting in consistent and well-constrained parameters. 

We note that \texttt{GALFITM} was only run on the optical images, not on the \textit{WISE} ones. This is because our methodology (described in \citealt{Buzzo_22b,Jarrett_12,Jarrett_13,Jarrett_19}) was already shown to provide reliable total magnitudes for the \textit{WISE} imaging.

For all galaxies, a single S\'ersic model was fitted in the optical bands, using as initial guesses the morphological parameters obtained by \cite{Poulain_21}. In our \texttt{GALFITM} setup, the magnitudes were allowed to vary freely amongst bands. Alternatively, the effective radius ($R_{\rm e}$), S\'ersic index ($n$), axis ratio ($b/a$) and position angle (PA) were allowed to vary, with the condition that they are constant in all bands. A discussion on appropriate degrees of freedom to use for each free parameter in \texttt{GALFITM} is provided by \cite{Buzzo_21b}.

The sky value for each wavelength band was obtained from the image header.
To run \texttt{GALFITM}, we created synthetic PSFs of each image using the \texttt{PSFex} routine \citep{Bertin_02}. The parameters used to create such PSFs were obtained using the \texttt{Source Extractor} routine \citep[\texttt{SExtractor},][version 2.19.5]{Bertin_02}.
We masked all interlopers, including GCs and nuclear star clusters, using the segmentation image output of \texttt{SExtractor}. The routine was run in single-mode in each band. We defined a source detection minimum area of 3 pixels and a threshold of 3$\sigma$ above the background to detect sources. \texttt{SExtractor} was run with an automatic background fitting, with a background cell size of 64 pixels. 

For the masking process, we visually inspected every mask to ensure no star forming region was being masked as this could bias the final results to older stellar populations. To do this, we carefully looked into \textit{GALEX} data on every one of the galaxies to see if any star formation was detected. As previously mentioned, for none of the galaxies UV emission was detected, indicating that the process was not masking any star forming region. 

\begin{figure*}
    \centering
    \includegraphics[width=\textwidth]{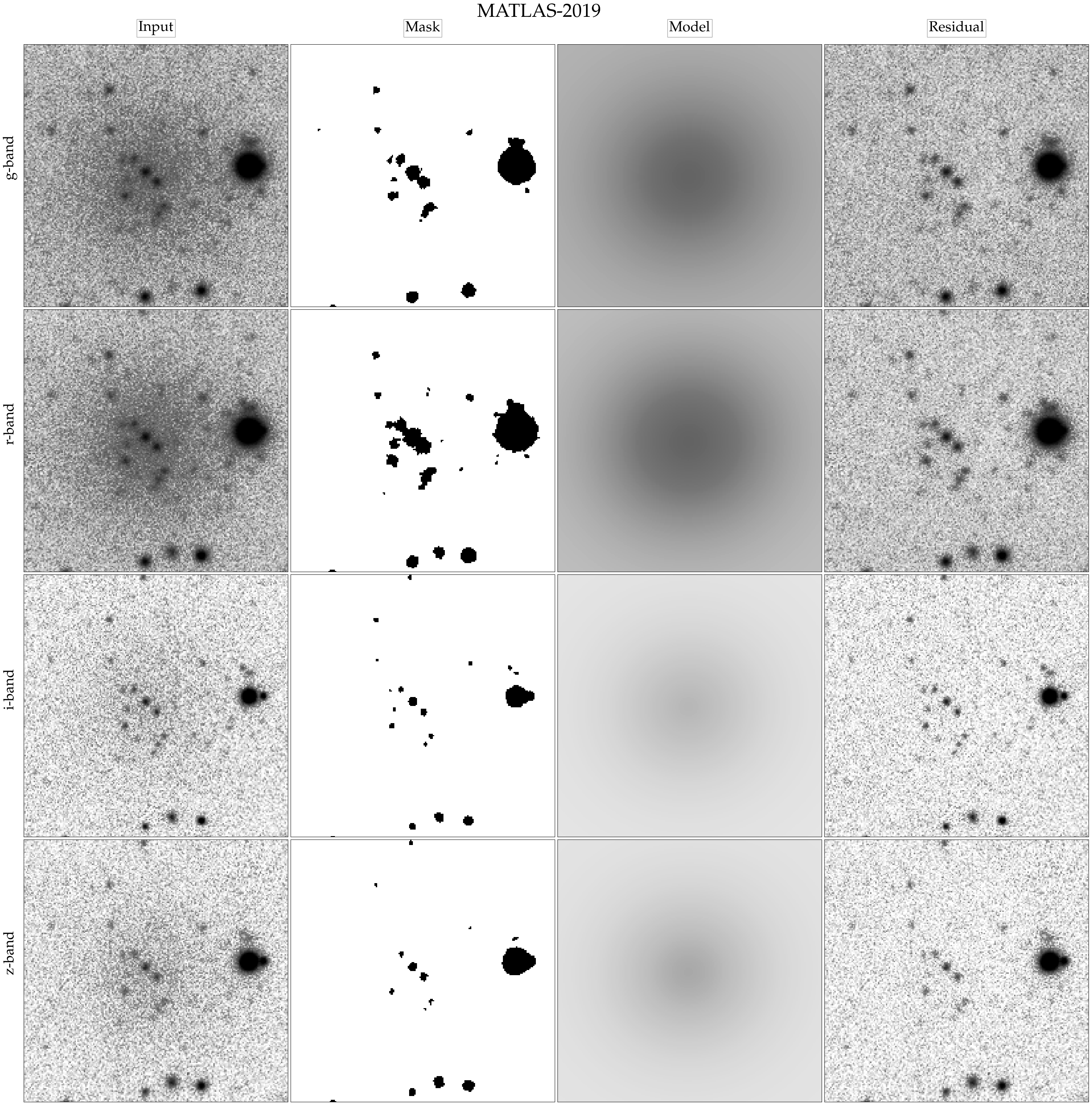}
    \caption{\texttt{GALFITM} model of MATLAS-2019 (NGC 5846\_UDG1). The first column shows the input image. Second column is the mask used in the model. The third column is the model. The fourth column is the residual (input -- model). Rows show the modelling in the $g$, $r$, $i$ and $z$ bands, respectively.}
    \label{fig:galfitm}
\end{figure*}

In Fig. \ref{fig:galfitm} we show a \texttt{GALFITM} model for the four optical bands of the galaxy MATLAS-2019 (NGC 5846\_UDG1). This example shows the highest mask fraction applied to our sample, as MATLAS-2019 is the galaxy with the most GCs within the sample of MATLAS UDGs. All other galaxies have had none to very few sources masked in front of them, therefore not compromising the final recovered photometry of the galaxies. \texttt{GALFITM} models for the other galaxies are of similar quality to the one shown in Fig. \ref{fig:galfitm}. 

Results from these fits are given in Table \ref{tab:morphology}.
We find that the MATLAS UDGs have a median S\'ersic index of $n = 0.95 \pm 0.14$, and a median axis ratio of $b/a = 0.62 \pm 0.13$.
\cite{Poulain_21} also obtained morphological parameters for all of the MATLAS dwarfs. Their results for the MATLAS UDGs show a median S\'ersic index of $n = 0.79 \pm 0.10$ and a median axis ratio of $b/a = 0.67 \pm 0.14$. Thus, their morphological findings for the MATLAS UDGs are consistent with ours.
\texttt{GALFITM}, similarly to \texttt{GALFIT}, provides unrealistically small uncertainties. In Appendix \ref{sec:appendix_galfit_unders}, we discuss some of the implications of using these underestimated uncertainties to the results obtained with SED fitting.

\subsection{\texttt{Prospector}}
\label{sec:prospector}

For the SED fitting, we run the fully Bayesian Markov Chain Monte Carlo (MCMC) based inference code \texttt{PROSPECTOR} \citep[][version 1.2.0\footnote{We note that tests carried out using \texttt{PROSPECTOR} version 1.0 have delivered systematically younger ages and longer star formation timescales.}]{Leja_17,Johnson_21}, complemented by the Flexible Stellar Population Synthesis package \citep[FSPS;][version 3.2]{Conroy_09,Conroy_10a,Conroy_10b}. To sample the posteriors, we used the dynamic nestled sampling \citep{Skilling_04,Higson_19} algorithm \texttt{dynesty} \citep{Speagle_20}.
\texttt{dynesty} was configured using 100 samples, 100 live points and a 0.05 tolerance when finishing the baseline run.

A full and thorough description of the main configuration and models used in \texttt{Prospector} is available in \citetalias{Buzzo_22b}. In the latter, \texttt{Prospector} models including dust attenuation provide results closer to those found with spectroscopy for UDGs. Thus, in this study, we only use models that include dust attenuation, assuming the \cite{Gordon_03} attenuation curve. Whether this dust is artificially added to improve the models or the galaxies actually have dust can only be probed using far infrared or ultraviolet deep data, which will be pursued in the future. We note, nonetheless, that tests with \texttt{Prospector} found a dust attenuation of $A_V \sim 0.1-0.2$ mag for Milky Way GCs when there should be none, suggesting an artificial addition of dust in \texttt{Prospector} models \citep[][ fig. 6]{Johnson_21}. 

For our fits of all galaxies, we include upper limit fluxes coming from the 12 and 22 $\mu$m bands from \textit{WISE}. For cases of extreme low S/N in the $z$-band of DECaLS, this band was also incorporated in the fitting as an upper-limit. For further understanding of how \texttt{Prospector} deals with upper limits, see Appendix A of \cite{Sawicki_12}. The inclusion of the upper limits was shown in \cite{Buzzo_22b} to be helpful in constraining the amount of dust attenuation found for these galaxies. In the cases where no dust is found, the upper limits do not play an important role in the fit.

We assume a delayed exponentially declining star formation history (SFH), as this is consistent with the recovered shape of the SFHs of UDGs by \citetalias{Ferre-Mateu_23}. The delayed model is an extension of the regular exponentially declining SFH. The latter usually assumes that star formation jumps from zero to its maximum value at the time $t_{\rm age}$ and then declines exponentially within a timescale $\tau$. The delayed model, alternatively, multiplies the exponentially declining SFH by the time since $t_{\rm age}$. This is capable of removing the discontinuity in the SFH at $t_{\rm age}$ and the condition that star formation can only decline after that point. This results in a more flexible, robust and physical SFH \citep[see][for thorough discussions and comparisons between parametric star formation histories]{Carnall_18,Leja_19}. 
Thus, the form of the chosen SFH is:

\begin{equation}
{\rm SFR } (t) \propto 
\begin{cases}
(t\,-\,t_{\rm age})\,\exp\left(-\frac{t - t_{\rm age}}{\tau}\right) {\rm , if \,\,} t\, >\, t_{\rm age} \\
0 {\rm , if \,\,  } t\, <\, t_{\rm age} \\
\end{cases}
\label{eq:sfh}
\end{equation}

According to this definition, $t_{\rm age}$ measures the onset of star formation within a galaxy and $\tau$ is an approximation of the star formation timescale, i.e., how long does a galaxy take to quench after reaching peak star formation. In this study, we are interested in the mass-weighted ages ($t_M$) for the galaxies, which are calculated analytically within \texttt{Prospector} using as input parameters the age since the onset of star formation ($t_{\rm age}$) and the star formation timescale ($\tau$). We note that \citetalias{Buzzo_22b} has incorrectly reported $t_{\rm age}$ as the mass-weighted age, not the true $t_M$ calculated analytically with \texttt{Prospector}. $t_M$ is only strongly different from $t_{\rm age}$ in the cases where $\tau$ is long, which is usually not the case for UDGs. Thus, the bulk of the comparisons between $t_{\rm age}$ from \citetalias{Buzzo_22b} and spectroscopic mass-weighted ages from the literature hold. However, some individual cases may have larger differences and be thus less consistent with spectroscopy than reported. Buzzo et al. (in prep.) is working on combining the dataset in this study with the one from \citetalias{Buzzo_22b}, where the photometry for all galaxies will be obtained consistently with \texttt{GALFITM} and the true $t_M$ will be reported for every galaxy so that we have a homogeneous dataset, where fair comparisons can be made and conclusions can be assessed.

For this study, we use two different configurations in \texttt{Prospector}. 

\begin{enumerate}
    \item $A_V \neq 0; z = z_{\rm host}$: five free parameters. Stellar mass (log $M_{\star}$/$M_{\odot}$), total metallicity ([M/H]\footnote{We use [M/H] rather than [Z/H] (as used in \citetalias{Buzzo_22b}), because Z denotes a mass fraction, and therefore [M/H] is less confusing.}), age ($t_{\rm age}$), star formation timescale ($\tau$) and diffuse interstellar dust ($A_V$). Redshifts ($z$) are fixed to the redshift of the group/massive galaxy where they were identified. The redshifts used come from \cite{Marleau_21}.\\
    
    \item $A_V \neq 0; z \neq z_{\rm host}$: six free parameters (log($M_{\star}$/$M_{\odot}$), [M/H], $t_{\rm age}$, $\tau$, $A_{V}$ and $z$). In this case, we leave the redshifts of the UDGs free.\\
\end{enumerate}

For these two scenarios, we placed linearly uniform priors on our free parameters. The range of the priors in our models are determined by the coverage of the Padova isochrones used in FSPS \citep{Marigo_07,Marigo_08}. These are:  log($M_{\star}$/$M_{\odot}$) = $6$ -- $10$, [M/H] = $-$2.0 to 0.2 dex, $\tau$ = 0.1--10 Gyr, $t_{\rm age}$ = 0.1--14 Gyr, $A_V = 0 - 4.344$ mag, and redshift $z =$ 0--0.045. This redshift range translates to a luminosity distance range of $0 < D_L < 200$ Mpc. Since all of the galaxies were identified in groups out to 45 Mpc, we assume that this redshift range is representative of the distance of all UDGs. We note that the default stellar masses obtained from \texttt{Prospector} are the total masses formed in the galaxy throughout its life. These were converted to surviving masses (most commonly quoted in the literature) in this work using a builtin function within \texttt{Prospector}.

As mentioned in \citetalias{Buzzo_22b}, we emphasize that different prior assumptions do not significantly alter our results or conclusions, with the exception of the prior assumption on the shape of the SFH, as discussed by \cite{Webb_22}.

\section{Results}
\label{sec:results}

\begin{figure*}
    \centering
    \includegraphics[width=\textwidth]{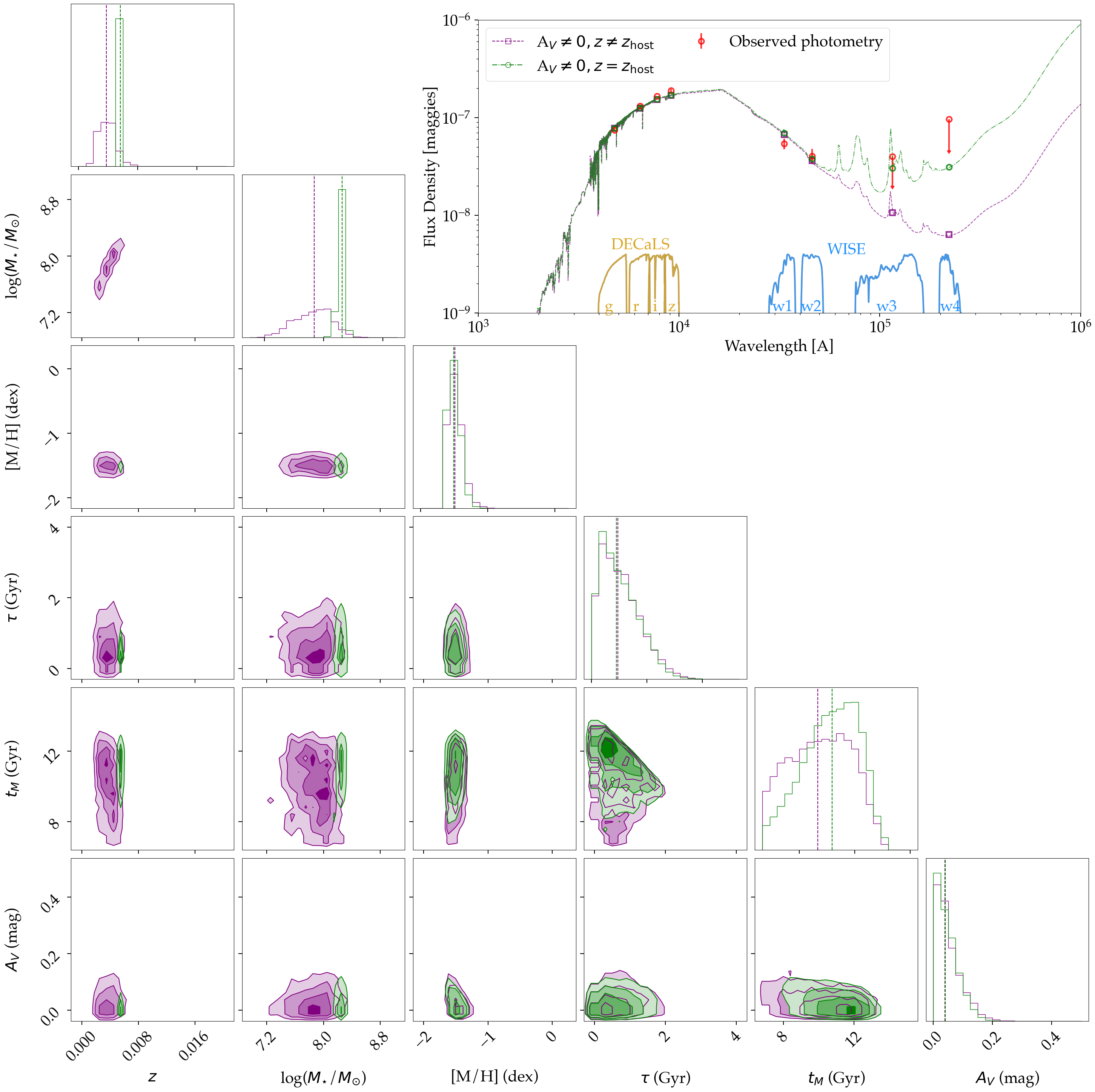}
    \caption{\texttt{Prospector} best-fit SED and covariance matrix for MATLAS-2019. In all panels, purple colours stand for the models with the redshift free ($z \neq z_{\rm host}$) and green colours are the models with the redshift fixed to the host ($z = z_{\rm host}$). \textit{Top-right} Best-fit spectra of MATLAS-2019 in both configurations. Red points are the observed magnitudes of the galaxy in the \textit{g}, \textit{r}, \textit{i}, \textit{z}, W1, W2, W3 and W4 bands, respectively. Yellow curves show the DECaLS transmission curves. Blue curves are the \textit{WISE} transmission curves. The remaining panels show the Monte Carlo Markov Chain (MCMC) corner plot comparing the posterior distribution of the two configurations. The first panel in each column shows the 1D posterior distribution of the fitted parameter, while the remaining panels show the correlation between parameters. This image can be read and interpreted as a covariance matrix. Columns stand for redshift, stellar mass, metallicity, star formation timescale, mass-weighted age and interstellar diffuse dust extinction. Metallicity, star formation timescale, age and dust attenuation are similar in both configurations. However, the stellar mass is strongly affected by the freedom of the redshift, as these two parameters are highly degenerate.}
    \label{fig:prospector_results_MATLAS-2019}
\end{figure*}

Here we present our results on the stellar population properties of the 59 MATLAS UDGs studied in this work. As explained in Section \ref{sec:prospector}, we carried out two model configurations in \texttt{Prospector}, one where the redshift is a free parameter and one with the redshift fixed. In this Section we show our results for both configurations, discuss the differences between them and the implications of using one or the other.

The resulting SEDs from both configurations of every galaxy were inspected to ensure that every photometric point was well modelled and that the shape was as expected. We found a median $\chi^2_{\rm reduced}$ of $1.41 \pm 2.65$. The best fits are for MATLAS-2103 and MATLAS-585, both with $\chi^2_{\rm reduced}$ = 1.01 and the worst is for MATLAS-1589, with $\chi^2_{\rm reduced}$ = 7.34.  Fig. \ref{fig:prospector_results_MATLAS-2019} shows a typical fit of an UDG in our sample, with $\chi^2_{\rm reduced}$ = 1.32. Upper limits were not included in the calculation of the $\chi^2$. Figs. \ref{fig:comparison_agemet} and \ref{fig:comparison_props} show the distribution of the parameters for all of the galaxies in our sample in both configurations. 

It is important to notice that choosing a parametric SFH for the galaxies comes with many limitations. One of such is that the star formation histories of the galaxies may have multiple components, i.e., an old quiescent and another with a recent burst of star formation. In such cases, a model with a parametric SFH that smooths over two distinct star formation episodes will be the most statistically correct, but will not reflect the real stellar populations present in the galaxies. Fitting the galaxies with a non-parametric SFH may be a solution for this problem, but it adds another one: too many free parameters for a limited amount of data, likely resulting in overfitting. To try and avoid this type of problem, we have gathered \textit{GALEX} data for all of the MATLAS UDGs to be able to correctly fit the star forming regions if the galaxies had any. None of the galaxies in our sample were detected in the FUV or NUV bands of \textit{GALEX}, and provided only very faint upper limits. We have tested including these UV upper limits on a subsample of our galaxies to understand the effect they would have on the recovered stellar populations. For all galaxies, the results are consistent with the ones obtained without the usage of \textit{GALEX}, indicating that \texttt{Prospector} is correctly tracing the dominant stellar populations within the galaxies only by using the optical to mid-IR data.

\subsection{Comparison between models with free and fixed redshifts}
\label{sec:redshifts}

\begin{figure*}
    \includegraphics[width=\textwidth]{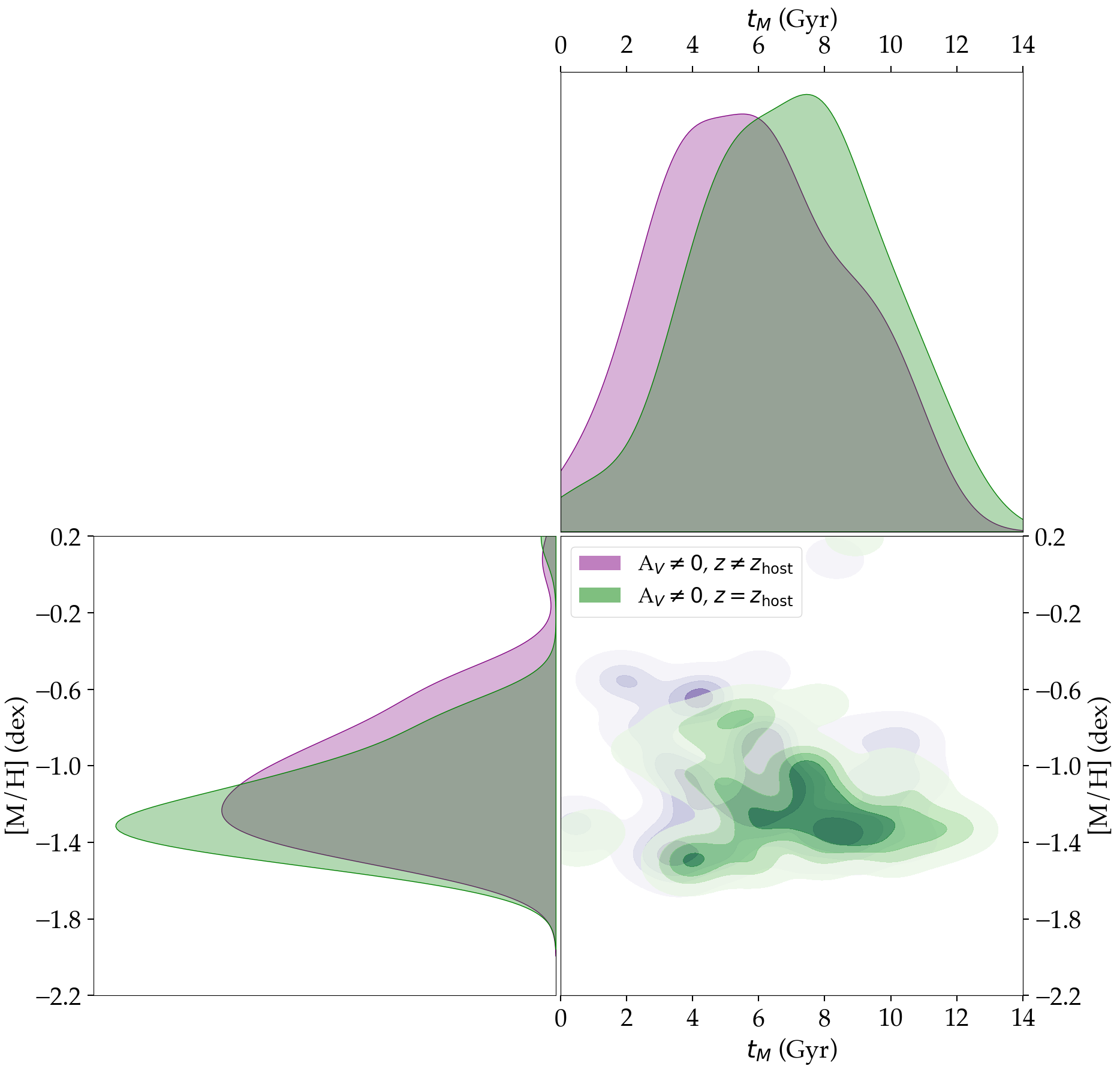}
    \caption{Age and metallicity of the 59 MATLAS UDGs recovered with \texttt{Prospector} using the two different configurations discussed in Section \ref{sec:prospector}. In all panels, purple colours stand for the models with the redshift free ($z \neq z_{\rm host}$) and green colours are the models with the redshift fixed to the host ($z = z_{\rm host}$). The bottom-right panel shows the age--metallicity distribution of the 59 MATLAS UDGs in both configurations. Histograms at the top and on the left show the marginal distributions of the data displayed on the age--metallicity plane. We conclude that models with the redshift free (purple) recover on average younger ages and more metal-rich populations for the UDGs.}
    \label{fig:comparison_agemet}
\end{figure*}

\begin{figure*}
    \includegraphics[width=\textwidth]{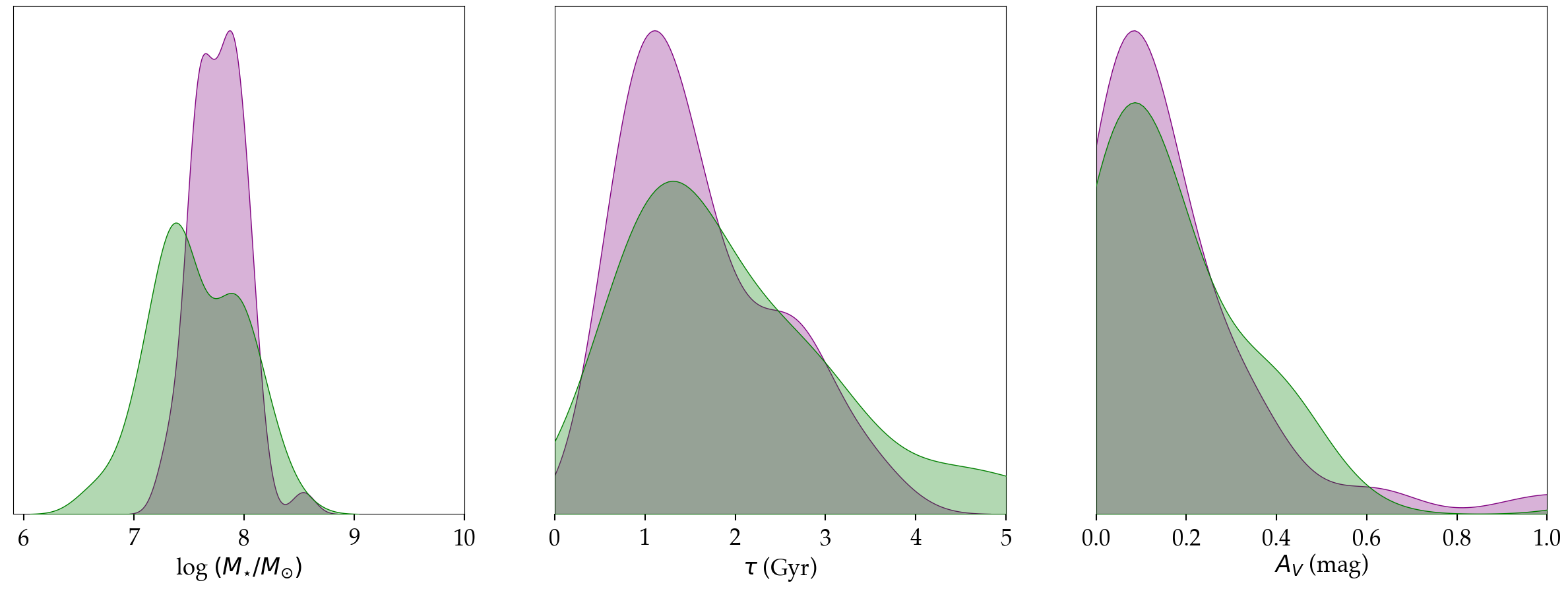}
    \caption{Stellar mass, star formation timescale and dust attenuation of the 59 MATLAS UDGs recovered with \texttt{Prospector} using the two different configurations discussed in Section \ref{sec:prospector}. In all panels, purple colours stand for the models with the redshift free and green colours are the models with the redshift fixed to the host. We conclude that models with the redshift free (purple) result in slightly more massive UDGs and a much narrower distribution centred at $\log (M_{\star}/M_{\odot}) \sim 7.8$ than models with the redshift fixed. No significant difference is observed in the star formation timescales and dust attenuation recovered in both configurations.}
    \label{fig:comparison_props}
\end{figure*}

\begin{figure}
    \includegraphics[width=\columnwidth]{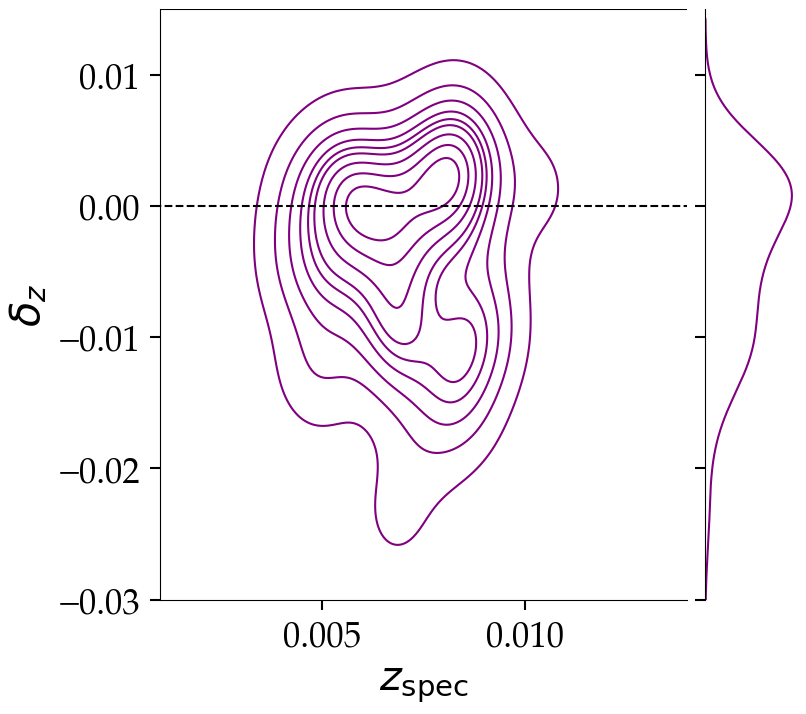}
    \caption{Accuracy in recovering photometric redshifts as a function of redshift. $\delta_z$ measures the difference between the spectroscopic redshifts of the candidate hosts and the recovered photometric redshifts of the UDGs. Marginal histogram shows the distribution of $\delta_z$. Positive values mean that UDGs were found to be in the foreground of their hosts. Negative values mean that UDGs were found in the background of their hosts. From previous findings \protect\citep{Heesters_23}, the expectation is that at least 75\% of the dwarf galaxies in the MATLAS survey are at the same redshifts as their hosts. Our results show a much larger portion of the UDGs at different redshifts ($> 70$\% of UDGs with |$\delta_z$|$> 0.0067$, equivalent to a difference in velocity of |$\delta_V$|$> 200$ km s$^{-1}$), indicating that photometric redshifts are not reliable at non-cosmological distances.}
    \label{fig:comparison_redshift}
\end{figure}

In Fig. \ref{fig:prospector_results_MATLAS-2019}, we show a comparison between the output best-fit SED and covariance matrix for the two configurations of \texttt{Prospector} applied to one of the galaxies in our sample, MATLAS-2019. 
It can be seen that the model where the redshift is free has a much larger spread in the determination of the stellar mass, while the results with the redshift fixed seem to be better constrained. Apart from the stellar mass, the other parameters do not seem to be strongly affected by the freedom of the redshift.

The same behaviour can be observed if we look into the whole population of galaxies, instead of an individual case. Figs. \ref{fig:comparison_agemet} and \ref{fig:comparison_props} provide a comparison of all of the stellar population properties of the MATLAS UDGs obtained with both configurations. As can be seen in these figures, the metallicity, star formation timescale, age and dust attenuation are not strongly affected by the freedom of fitting or lack thereof to fit redshift in the models. Galaxies are found to be on average younger and more metal-rich in the models where the redshift is fixed, but consistent within the uncertainties with the stellar populations found in the models with the redshift free. However, the stellar mass of the galaxies shows a larger difference when comparing models. Overall, galaxies have higher stellar masses in the models where the redshift is a free parameter. That is because in integrated SEDs at non-cosmological distances, the redshift and stellar mass are highly degenerate parameters.

As explained in Section \ref{sec:data}, we expect that most UDGs in our sample are at the same redshift as their hosts in the models with the redshift free, if our redshift estimates are reliable. We explore the difference between the redshift of the host and our recovered photometric redshift ($\delta_z$) in Fig. \ref{fig:comparison_redshift}.
As can be seen, while there is a peak at $\delta_z = 0$, a significant portion ($> 70$\%) of the UDGs have |$\delta_z$|$> 0.0067$, equivalent to a difference in velocity of |$\delta_V$|$> 200$ km s$^{-1}$). Thus, they do not have photometric redshifts consistent with those of their hosts. Most of them are in fact found in their background (i.e., $\delta_z <0$). Since photometric redshifts at non-cosmological distances are very uncertain, we consider $\delta_z$ being different than zero as a confirmation that redshifts with this method cannot be trusted rather than an indication that MATLAS UDGs are not at the redshifts of their hosts. 

Going forward, we treat the models with fixed redshift as the best representation of our galaxies. Thus, from now on, we will only discuss the results of this configuration. 
We reinforce, nonetheless, that even if the galaxies are not at this assumed distance, changes in the stellar populations (except stellar mass) between models are small and well within their uncertainties (see Figs. \ref{fig:prospector_results_MATLAS-2019}, \ref{fig:comparison_agemet}, and \ref{fig:comparison_props}). 
We carefully checked that the changes in stellar mass do not strongly affect any of our conclusions that follow. Specifically, we have checked that the mass--metallicity bimodality found for the models with fixed redshift and thoroughly discussed in Section \ref{sec:MZR} is also present in the models where the redshift is a free parameter. We have also checked that the change in stellar mass is small and not enough to change the classification of the UDGs according to their positioning on the mass--metallicity plane (i.e., which MZR they follow).

\subsection{Median stellar populations}

The results of the models with the redshift fixed at the redshift of the host galaxy are fully presented in Table \ref{tab:stellarpops}. The median and median absolute deviations of the stellar populations of the whole sample of MATLAS UDGs are presented below.
We find that the MATLAS UDGs have intermediate-to-old ages, with a median mass-weighted age of $t_M = 7.1 \pm 1.8$ Gyr. The galaxies display an average metal-poor population with [M/H] = $-1.2 \pm 0.2$ dex. These stellar populations are equivalent to those found for UDGs with spectroscopy by \citetalias{Ferre-Mateu_23}. We find an average $\tau$ of $1.6 \pm 0.7$ Gyr. Finally, the median interstellar diffuse dust attenuation coming from the SED fitting of the galaxies in our sample is $A_{V} = 0.12 \pm 0.07$ mag, consistent with the expectation that these galaxies should have little-to-no dust component and with the caveat that the non-zero dust content could be a small artificial addition that compensates for inaccuracies in the SED models. It is interesting to notice that although our dust priors extend out to 4.3 mag, the highest $A_V$ value found is 1.17 mag, highlighting the importance of the inclusion of the \textit{WISE} upper limits from the 12 and 22$\mu$m bands to constrain the amount of dust in the galaxies. This power of the near- and mid-infrared bands in constraining the dust amount found had been discussed previously both by \cite{Pandya_18} and \citetalias{Buzzo_22b}. Nonetheless, as expected, UDGs with higher amounts of dust show higher degrees of degeneracy between the dust extinction and the metallicity and age.

Trends between the recovered stellar populations and the galaxy morphologies, environments, scaling relations and GC-richnesses are investigated in Section \ref{sec:MZR}.

\subsection{Comparison with the literature}
\label{sec:comp_literature}

In this Section, we compare our SED fitting results with stellar populations recovered with spectroscopy in the literature for some of the MATLAS UDGs.

\begin{itemize}
\item \textbf{MATLAS-2019}\\

MATLAS-2019, also known as NGC 5846\_UDG1 \citep{Forbes_21}, has been heavily studied in the past years due to its unusually large GC population \citep[][Haacke et al. in prep.]{Mueller_20,Mueller_21,Danieli_22}. 
In terms of its stellar populations, MATLAS-2019 has been spectroscopically studied by \cite{Mueller_20}, \cite{Heesters_23} and \citetalias{Ferre-Mateu_23}.

\cite{Mueller_20} found, using VLT/MUSE data, that MATLAS-2019 has a metallicity of [M/H] = [Fe/H] = $-1.33^{+ 0.19}_{- 0.01}$ dex, an age of $11.2^{+ 1.8}_{- 0.8}$ Gyr and a stellar mass of $3.6 \times 10^8 \,{\rm M}_{\odot} \,(\log \,(M_{\star}/M_{\odot}) = 8.56)$. Morphological parameters for MATLAS-2019 were also obtained by \cite{Mueller_20}, who found a S\'ersic index of $n = 0.73 \pm 0.01$, an axis ratio of $b/a = 0.90 \pm 0.01$ and an effective radius of $R_{\rm e} = 17.2$ arcsec ($\sim 2.2$ kpc at the assumed distance of 26.3 Mpc).
\cite{Heesters_23} studied MATLAS-2019 using the exact same VLT/MUSE data as \cite{Mueller_20}. They found, nonetheless, much older and more metal-poor populations, i.e., [M/H] = [Fe/H] = $-1.88^{+ 0.13}_{- 0.06}$ dex and an age of $13.5^{+ 0.5}_{- 0.2}$ Gyr. A reason for this discrepancy is not discussed by \cite{Heesters_23}.
Finally, \citetalias{Ferre-Mateu_23} has studied MATLAS-2019 using Keck/KCWI data. They found a metallicity of [M/H] = $-1.15 \pm 0.25$ dex, an age of $8.2 \pm 3.05$ Gyr and a stellar mass of $\log(M_{\star}/M_{\odot}) = 8.1$, assuming a distance of 24.89 Mpc to the UDG \citep{Forbes_21}. The cause for this difference is yet to be understood, however, we note that the S/N of the spectra used by \citetalias{Ferre-Mateu_23} is higher (S/N $\sim$ 15-20 pix$^{-1}$) than that used by both \cite{Mueller_20} and \cite{Heesters_23} (S/N = 12.4 pix$^{-1}$). 

With \texttt{Prospector}, we recover a metallicity of [M/H] = $-1.40^{+ 0.10}_{- 0.12}$ dex, a mass-weighted age of $11.22^{+ 1.79}_{-3.23}$ Gyr and a stellar mass of $\log (M_{\star}/M_{\odot}) = 8.01^{+ 0.03}_{- 0.05}$ for MATLAS-2019, for a distance of 25.2 Mpc (i.e., the distance to the host NGC 5846). Our study is consistent within the uncertainties with both \cite{Mueller_20} and \citetalias{Ferre-Mateu_23}. The results, nonetheless, in terms of age and metallicity, are closer to the ones found by \cite{Mueller_20}, indicating a prevalence of a metal-poor and old population.
We found MATLAS-2019 to be younger and more metal-rich than the results obtained by \cite{Heesters_23}.
Additionally, the morphological parameters obtained with \texttt{GALFITM} for MATLAS-2019 were $n= 0.6 \pm 0.1$ , $b/a = 0.97 \pm 0.03$ and $R_{\rm e} = 15.7 \pm 0.4$ arcsec. These are all consistent within the uncertainties with the findings of \cite{Mueller_20}. \\

\item \textbf{MATLAS-585 and MATLAS-2103}\\

These two MATLAS UDGs were recently studied using VLT/MUSE data by \cite{Heesters_23}.
MATLAS-585 was found to have a metallicity of [M/H] = [Fe/H] = $-1.88^{+ 0.17}_{- 0.13}$ dex and an age of $9.0^{+ 2.9}_{- 3.1}$ Gyr. % and a M/L = $1.1^{+ 0.7}_{- 0.0}$.
With \texttt{Prospector}, we find that MATLAS-585 has a metallicity of [M/H] = $-1.36^{+ 0.27}_{- 0.16}$ dex and an age of $8.91^{+3.44}_{-4.31}$ Gyr. Thus, our results in terms of age are consistent with those of \cite{Heesters_23}, but they find a much more metal-poor population for MATLAS-585 than we do with \texttt{Prospector}.

MATLAS-2103 was found by \cite{Heesters_23} to have a metallicity of [M/H] = [Fe/H] = $-1.69^{+ 0.01}_{- 0.22}$ dex and an age of $11.3^{+ 2.2}_{- 2.5}$ Gyr. % and a M/L = $1.5^{+ 0.6}_{- 0.1}$.
In this study, we find that MATLAS-2103 has a metallicity of [M/H] = $-1.40^{+ 0.14}_{- 0.13}$ dex and an age of $9.79^{+ 2.81}_{- 3.84}$ Gyr. Similar to the case of MATLAS-585, we find ages consistent with those of \cite{Heesters_23}, but their stellar populations are more metal-poor than ours.
Notably, \cite{Heesters_23} tends to find systematically more metal-poor stellar populations for galaxies if we take this study, \citetalias{Ferre-Mateu_23} and \cite{Mueller_20} as baselines.
\\

\end{itemize}

Overall, we show that our results are consistent with those found using spectroscopy. Our SED fitting technique had already been shown to provide consistent results with spectroscopy in the studies of \citetalias{Buzzo_22b} and \citetalias{Ferre-Mateu_23}. In both cases our results were shown to have a metallicity offset of $\sim$ 0.2 dex, which is well within our uncertainties. We reinforce that while SED fitting results have much larger uncertainties, stellar populations obtained from broad-band photometry are far less restricted than spectroscopic observations in depth and spatial coverage. Additionally, the ability of this technique to gather information from a wide wavelength range makes up for the loss of detailed features provided by spectroscopic data. SED fitting also allows for complete sample studies such as this one, which are prohibitively expensive to do spectroscopically on big telescopes. Apart from the larger uncertainties, another downside of SED fitting is the difficulty in obtaining [$\alpha$/Fe] and dynamical information from the galaxies. Thus, spectral and photometric information are complementary and should be jointly used to better understand UDGs.

\section{Discussion}
\label{sec:discussion}

As mentioned in the Introduction, the stellar populations of UDGs can help understanding their formation histories. This can be made by means of comparing these populations to those of different classes of galaxies to try and identify similarities/differences between them. In Section \ref{sec:MZR}, we use scaling relations (i.e., the mass--metallicity relation) to compare UDGs to classical dwarfs (observations) and high-redshift galaxies (simulations) to try and elucidate their origins. In Section \ref{sec:two_types} we discuss if there is evidence for two classes of UDG based on their stellar populations and morphological properties and associate these classes with their most likely formation scenarios.

\subsection{Mass--metallicity relation: Clues to the origins of UDGs}
\label{sec:MZR}

We explore in this section the positioning of MATLAS UDGs on the stellar mass -- metallicity plane, especially in comparison with known mass -- metallicity relations (MZR). We investigate trends with different morphological and stellar population parameters, as well as GC--richness. 

\begin{figure}
    \centering
    \includegraphics[width=\columnwidth,trim=0 3cm 0 0, clip]{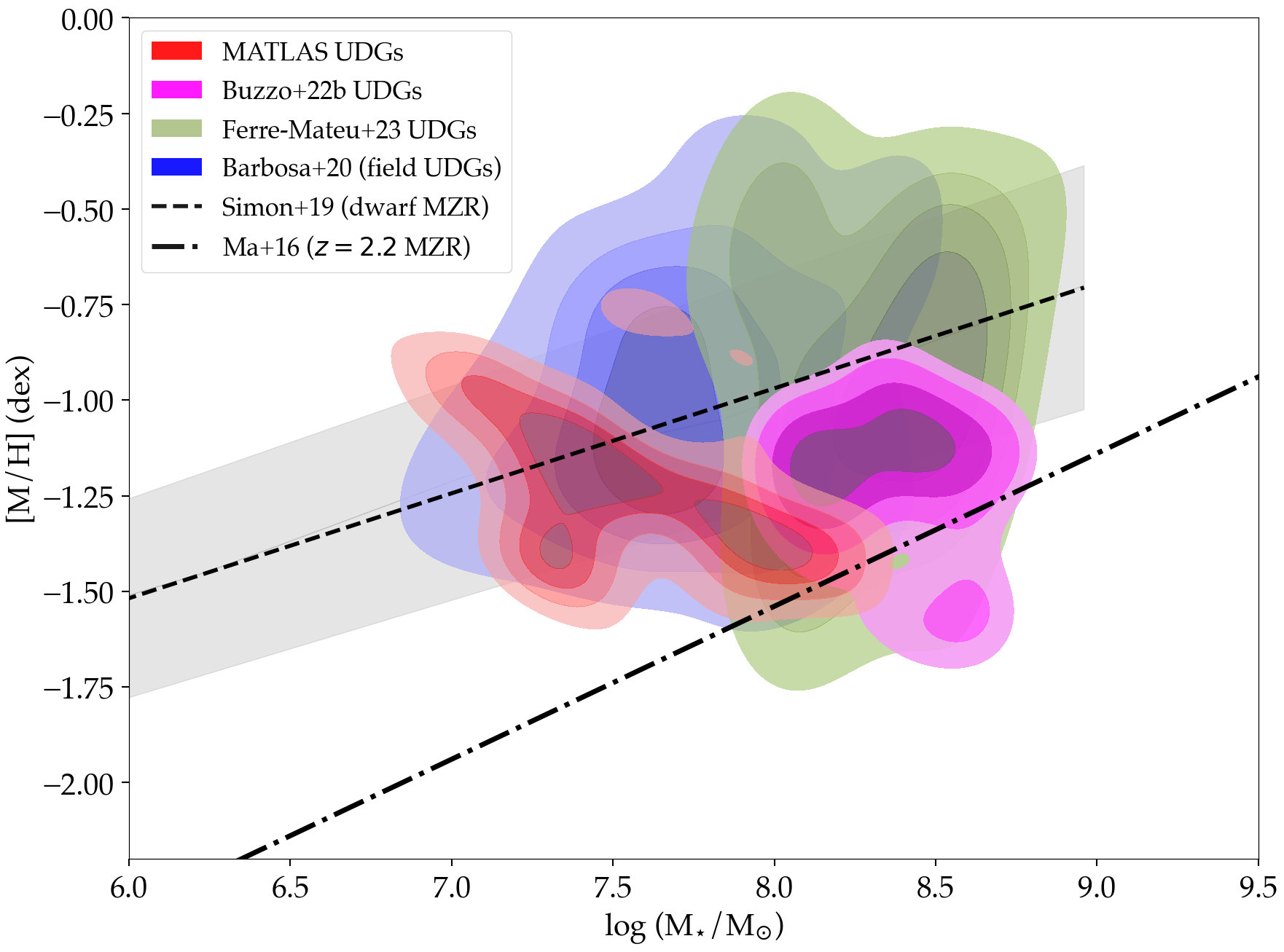}
    \includegraphics[width=\columnwidth]{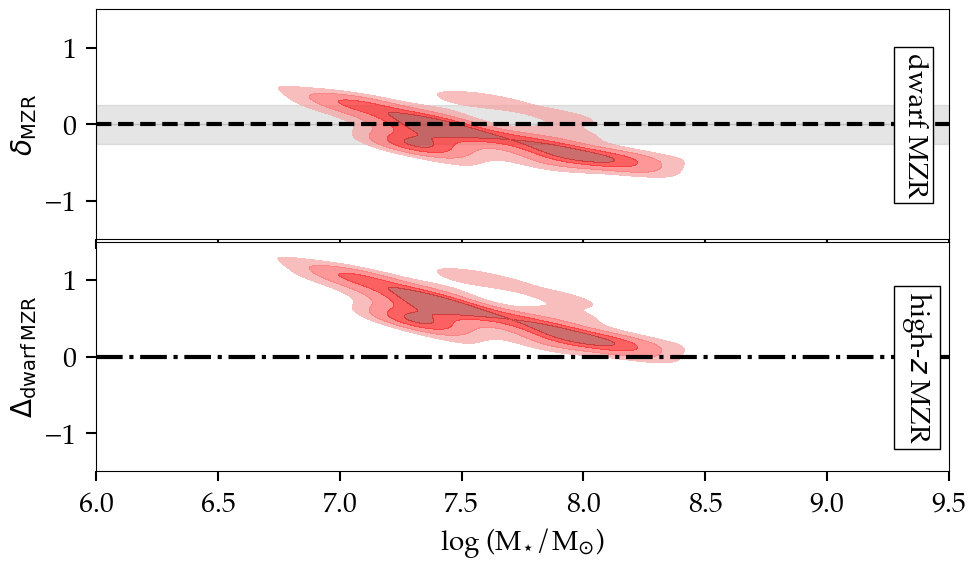}
    \caption{Stellar mass--metallicity distribution of UDGs. \textit{Top panel:} MATLAS UDGs are shown in red. Field UDGs from \protect\cite{Barbosa_20} are shown in blue. UDGs in different environments from the spectroscopic sample of \citetalias{Ferre-Mateu_23} are shown in green. UDGs in a variety of environments from \protect\citetalias{Buzzo_22b} are shown in magenta. The \protect\cite{Simon_19} MZR for classical Local Group dwarf galaxies is shown with the black dashed line. The dash-dotted line is the evolving MZR at redshift $z=2.2$ from \protect\cite{Ma_15}. \textit{Middle panel:} Difference in metallicity between the MATLAS UDGs and the \protect\cite{Simon_19} classical dwarf MZR. UDGs at $\delta_{\rm MZR} = 0$ follow perfectly the MZR, if $\delta_{\rm MZR} > 0$, then the UDG lies above the MZR, if $\delta_{\rm MZR} < 0$, then the UDG lies below the MZR. \textit{Bottom panel:} Difference in metallicity between the MATLAS UDGs and the \protect\cite{Ma_15} MZR at $z=2.2$. MATLAS UDGs show a bimodality in the mass--metallicity distribution, with the first mode being consistent with the classical dwarf MZR and the second being more consistent with the high-redshift MZR.}
    \label{fig:simple_MZR}
\end{figure}

For reference, in Figs. \ref{fig:simple_MZR} and \ref{fig:MZR_matlasonly}, we plot two MZRs: \cite{Simon_19} for classical dwarfs, and the simulated MZR from \cite{Ma_15} for evolving sources at a redshift of $z=2.2$. To be consistent, we plot total metallicities from \texttt{Prospector} on top of the converted to [M/H] \cite{Simon_19} relation, from its original in [Fe/H] \citep{Vazdekis_15}. A full discussion on how to make this conversion and the caveats that arise in doing so are given in \citetalias{Buzzo_22b}. In Fig. \ref{fig:simple_MZR} we show the distribution of the MATLAS UDGs in the mass--metallicity plane in comparison to three other works in the literature. In blue we show the study of 100 field UDGs using SED fitting from \cite{Barbosa_20}. The study of 29 UDGs in a variety of environments also using SED fitting from \citetalias{Buzzo_22b} is shown in magenta. Lastly, the largest spectroscopic study of UDGs to date (25) from \citetalias{Ferre-Mateu_23} is shown in green. The UDGs in \citetalias{Ferre-Mateu_23} are in multiple environments, but biased towards clusters.

The MATLAS UDGs, at first glance, seem to form a new MZR, with a metallicity that is anti-correlated with the stellar mass. However, the population shows a bimodality, indicating that instead of an anti-correlation, the MATLAS UDGs split into two samples. One of them is well within the classical dwarf MZR and another has lower metallicities and lies closer to the simulated high-$z$ MZR. To make this bimodality clearer, in the bottom panels of Fig, \ref{fig:simple_MZR}, we show the residual MZR (i.e., the difference between the metallicities of the UDG population and the value expected from the MZR). It can be seen that the low-mass metal-rich UDGs are well within the scatter of the classical dwarf MZR, while the high-mass metal-poor UDGs are further down and in fact are more consistent with the MZR at high-redshift. This bimodality may be indicative of different formation scenarios for the UDGs populating each mode.

When comparing to the works in the literature, we see that while all have a population of UDGs that follows well the classical dwarf MZR, they all seem to probe different regions of the parameter space. \cite{Barbosa_20}, as well as the MATLAS UDGs, are biased towards lower mass UDGs in low density environments, while \citetalias{Buzzo_22b} and \citetalias{Ferre-Mateu_23} are biased towards more massive cluster UDGs. \citetalias{Ferre-Mateu_23}, in addition, probes a population of more massive, metal-rich, cluster UDGs, although overall their distribution scatters around the MZR relations at log($M_{\star}/M_{\odot}$) $\sim 8$.

\cite{Barbosa_20} find many UDGs to be more metal-rich than the classical dwarf MZR. These were found to be the youngest in their sample. 
\citetalias{Ferre-Mateu_23} found a similar subpopulation of UDGs more metal-rich than the classical dwarf MZR, as well as another subpopulation that is more metal-poor. We recover the population of metal-poor UDGs found by them, but no UDGs in the same mass range that are metal-rich. The explanation for this is that the sample of \citetalias{Ferre-Mateu_23} is biased towards the cluster environment where tidal effects are more common \citep{Sales_22} and result in the higher metallicities found. 

Finally, the sample covered by \citetalias{Buzzo_22b}, although more representative than the current sample in terms of environments, is more constrained in terms of stellar masses, showing only a small range around 8.0 $\leq$log($M_{\star}/M_{\odot}$) $\leq$ 8.5. Similar to the findings of \citetalias{Ferre-Mateu_23} and the ones of the current study, they found that a subsample of the UDGs follows the classical dwarf MZR, while another population is much more metal-poor and is more consistent with the high-$z$ MZR.

\subsubsection{Dependence with morphology and local environment}

\begin{figure*}
    \centering
    \includegraphics[width=\textwidth]{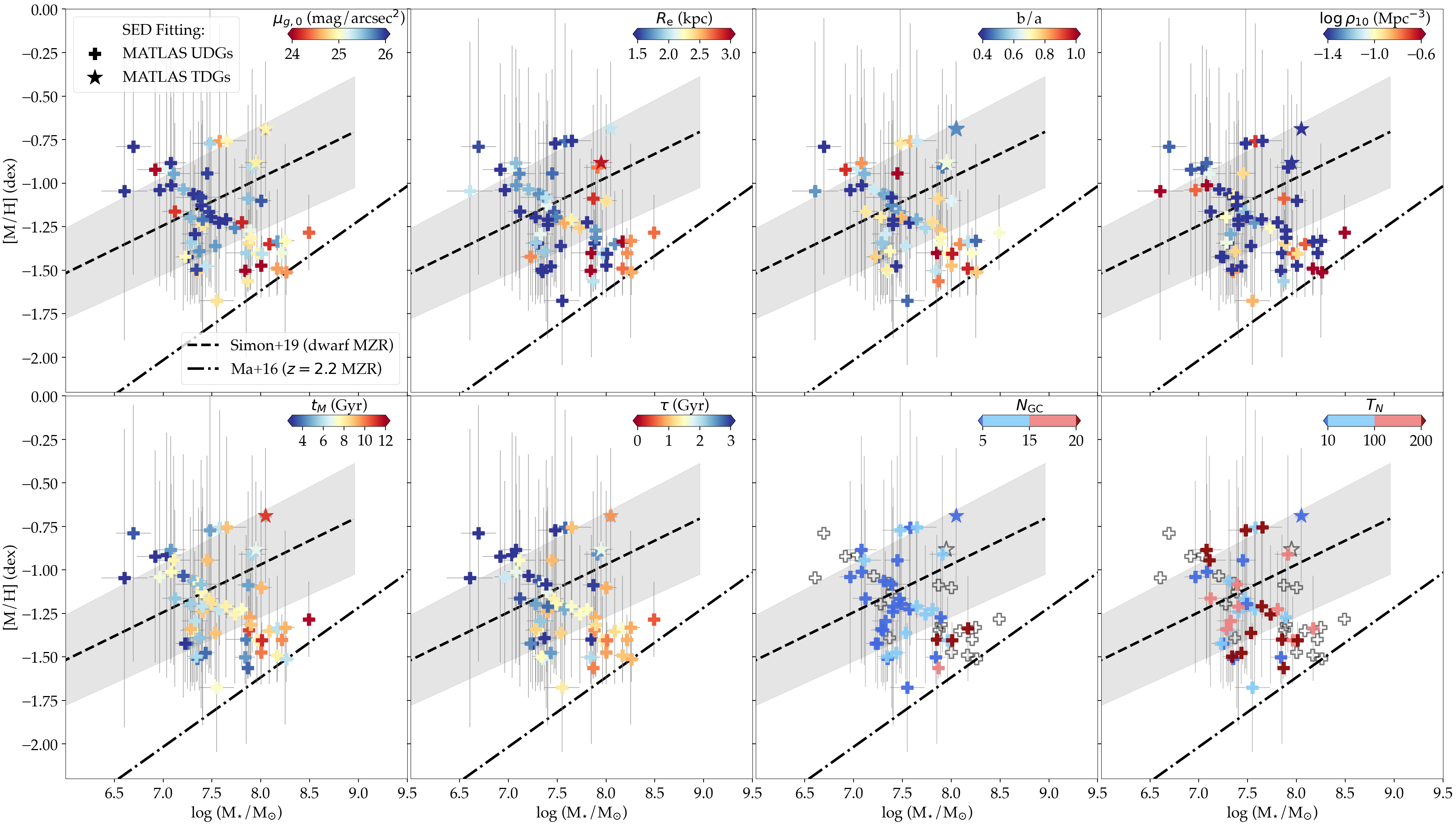}
    \caption{Distribution of the physical properties of the MATLAS UDGs in the stellar mass--metallicity plane. Plus signs show the results obtained from SED fitting with \texttt{PROSPECTOR} for the 59 UDGs in our sample, with the 2 confirmed tidal UDGs in our sample being highlighted with star symbols (MATLAS-1824 and MATLAS-478). \textit{Top-row:} UDGs are colour-coded by surface brightness, effective radii, axis ratio and local environment \protect\citep{Marleau_21}, respectively. \textit{Bottom-row:} UDGs are colour-coded by their mass-weighted ages, star formation timescales, number of GCs (Marleau et al. subm.) and GC specific frequencies (per unit stellar mass), respectively. The \protect\cite{Simon_19} MZR for classical Local Group dwarf galaxies is shown with the dashed black line. The dash-dotted black line is the evolving MZR at redshift $z=2.2$ from \protect\cite{Ma_15}. An alternative view of this figure is shown in Appendix \ref{fig:direct_plots}. These results indicate a possible bimodality in the MATLAS UDGs population: Population i) UDGs are fainter, smaller, elongated, live in less dense environments, are younger, have prolonged SFHs, have a smaller number of GCs and on average smaller specific frequencies. Population ii) UDGs in this population are brighter, bigger, rounder, live in denser environments, are older, have shorter SFHs, host on average more GCs and have on average larger specific frequencies. }
    \label{fig:MZR_matlasonly}
\end{figure*}

The fact that the morphology of galaxies is correlated with the environment they reside in has been known since the 1970's, giving origin to the so-called `Morphology--Density Relation' \citep{Oemler_74,Davis_76,Dressler_80,Dressler_97}, first proposed by Edwin Hubble in the 1930's \citep{Hubble_31}. In a very simplistic explanation, early-type galaxies are found to occupy the densest environments such as the centres of clusters of galaxies, while late-type galaxies are observed in their outskirts and to be prevalent in less dense environments, such as the field. The morphology--density relation was also found to hold for classical dwarf galaxies \citep{Phillipps_98, Penny_11}, showing that they behave in a similar way to their massive counterparts in terms of the environments they inhabit. Noteworthy, \cite{Poulain_21} also found a morphology--density relation for the dwarfs on the MATLAS fields comparable to that seen in dense clusters, and \cite{Jiaxuan_23} found a similar behaviour for UDGs surrounding Milky Way analogs.

Various works in the literature have looked into the morphological parameters of UDGs \citep[e.g.,][]{Rong_Yu_20,KadoFong_20,KadoFong_21,vanNest_22,Jiaxuan_23}, such as their ellipticities, which can hold much information about the physical processes involved in forming UDGs. \cite{Rong_Yu_20}, for example, found that the morphologies of cluster UDGs depend on their luminosities, environments and redshifts. In terms of axis ratios, they showed that there are at least two populations of UDGs, the elongated ($b/a \sim$ 0.4) and the round ones ($b/a \sim$ 0.9). In their study, UDGs at higher redshifts ($z \sim 0.3$) are more elongated than their low-redshift counterparts, indicating a possible evolutionary path of UDGs becoming rounder with time. They also show that bright ($M_g > -15.2$ mag) UDGs are rounder than the fainter ones ($M_g < -15.2$ mag). Finally, they showed that UDGs closer to the centre of clusters are rounder than the ones in the outskirts or outside the cluster virial radius. Their findings apply to both nucleated and non-nucleated UDGs, although it is worth noting that the number of nucleated UDGs increases towards the centre of clusters and these are the roundest sources. A bimodality in the axis ratio, mass, and luminosity of UDGs has also been reported in the Coma and Virgo clusters by \cite{Lim_18} and \cite{Lim_20}, respectively. They find that UDGs that are dark matter dominated (M/L > 1000) have relatively rounder shapes (higher $b/a$), while UDGs with M/L $\sim$ 500 are more elongated (low $b/a$). They also found that the round UDGs have higher GC specific frequencies than the more elongated ones.
 
In the top row of Fig. \ref{fig:MZR_matlasonly}, we show the MATLAS UDGs colour-coded by their surface brightnesses, effective radii, axis ratio and local volume density. Similar to the findings of \cite{Rong_Yu_20}, we also find a correlation between the surface brightness and the mass of the galaxies. Less massive UDGs are, on average, the faintest ones, while more massive UDGs are the brightest. This is expected given the narrow range in surface brightness and size, and the fact that the S\'ersic index of the galaxies is close to 1. In addition, a similar correlation exists for the effective radii of the galaxies. Less massive UDGs are smaller, while the more massive ones are bigger. This behaviour is not unique to UDGs, it is in fact expected from the mass-size relation, but it is reassuring to confirm this expectation in our sample of galaxies.

In terms of their axis ratios, we can see a weak correlation with both the mass and metallicity, where more elongated UDGs seem to be positioned in a less massive/ more metal-rich region, while the roundest UDGs occupy a position corresponding to more massive, more metal-poor populations. 
There is a non-negligible portion of UDGs also with round morphologies located in the upper parts of the plane. One explanation for this is that the ellipticity of galaxies is highly dependent on their inclinations, which could drive oblate/flattened galaxies to look round if they are face-on. It is also worth noticing that more massive edge-on UDGs would likely be pushed over the surface brightness threshold definition for UDGs due to their higher apparent brightness, which can explain why we see such a correlation between the surface brightness, axis ratio and mass of the galaxies. Additionally, less massive galaxies tend to be rounder \citep[e.g., ][]{SanchezJanssen_19}, however, these faint galaxies are below our mass range, also explaining why we do not have a population of round galaxies at lower masses.

We note the study of \cite{Cardona_Barrero_20}, who used the NIHAO simulations to suggest a correlation between the morphology and kinematics of UDGs. They suggested that more elongated UDGs (i.e., smaller axis ratios) are rotationally-supported, while the rounder ones (higher axis ratios) are pressure-dominated. These findings are further supported for classical dwarfs by Pfeffer et al. (subm.).

If we now compare these findings with the panel where the UDGs are coloured according to the density of the environment they reside in (i.e., smaller values of $\log \rho_{10}$ are for galaxies that are fairly isolated, while higher values signify that galaxies are in more dense environments), we can start to see some forms of the morphology--density relation appearing. That is, UDGs that show the most elongated morphologies are the ones located in the less dense environments. Alternatively, UDGs that are rounder tend to be in the most dense environments.
If we consider that UDGs in denser environments are prone to more mergers and encounters than their isolated counterparts, then having rounder morphologies is a natural consequence of that \citep{Moore_96,Moore_98,Moore_99}. We caveat that this interpretation is done for the full sample of galaxies, and individual cases may differ from the ensemble properties.

If we analyse the environment panel on its own, there appears to be a weak trend between the environment and the positioning of the UDGs, where the majority of the UDGs in the most dense of our probed environments are below the classical dwarf MZR. We remind the reader that all of the MATLAS UDGs are, by definition, in low-to-moderate density environments, and thus, the range of densities probed is limited and does not include the most dense environments, such as clusters. Selection effects can be playing a role in our findings and, thus, comparisons with UDGs located in higher density environments are necessary and will be pursued in a follow-up study (Buzzo et al. in prep.). It is interesting, nonetheless, to see that even with such a small range of densities, a correlation between the environment and the stellar populations can still be weakly identified. 
Although a correlation was identified with the local volume density, we found no correlation between any properties and the projected groupcentric distance of the UDGs. 
We emphasise that these are the interpretations for the ensemble of MATLAS UDGs, but as can be seen, individual exceptions may exist (e.g., round UDGs at low-density environments or elongated ones at the highest probed densities).  
We use colour-coding to show the changes in the properties of the MATLAS UDGs across the mass--metallicity plane in Fig. \ref{fig:MZR_matlasonly}, but for an alternative view, we provide in Fig. \ref{fig:direct_plots} direct plots of each property against their positioning on the plane to emphasise the discussed trends.

\subsubsection{Dependence with age and star formation timescale}

We now turn to understanding how the positioning of the UDGs in the stellar mass -- metallicity plane depends on their star formation history properties, e.g., their ages and star formation timescales. 

\cite{Ferre-Mateu_18} have shown that there exists an age dependence with projected clustercentric distance, i.e., UDGs are younger at larger projected clustercentric radii. \cite{Alabi_18} and \cite{Kadowaki_21} also found a colour dependence with the environment, with bluer UDGs residing in lower-density environments than the redder ones. 
Finally, \citetalias{Buzzo_22b} found, using SED fitting, that UDGs in the field are on average younger and more metal-rich than cluster UDGs. Group UDGs in their work were found to be in a transitional space in the age--metallicity plane. However, these were based on very small number statistics, hence the need for this current study focused mainly on group UDGs. Similar to these recent findings, the ages of the MATLAS UDGs also seem to distill into two main types: younger UDGs following the dwarf MZR and older ones following the high-$z$ MZR. We however caution the reader that ages recovered with \texttt{Prospector} are very sensitive to the version of the code used (see Section \ref{sec:prospector}), as well as to the uncertainties in the optical photometry (see Appendix \ref{sec:appendix_galfit_unders}).

One interesting recent finding on the stellar populations of UDGs by \citetalias{Ferre-Mateu_23} is a correlation between the $\alpha$ enrichment ([$\alpha$/Fe]) and the star formation histories of UDGs. Using a phase-space diagram, which may be particularly related to the environment of the UDGs, they show that the UDGs with the highest [$\alpha$/Fe] are in the region corresponding to early infallers, while UDGs with the lowest [$\alpha$/Fe] tend to be in the outskirts of clusters or in less dense environments. The $\alpha$-enrichment can be associated with the speed at which star formation starts and quenches within galaxies. Thus, this parameter can very well separate early-quenched galaxies (high [$\alpha$/Fe]) from galaxies with prolonged star formation histories (low [$\alpha$/Fe]).

With SED fitting we cannot recover [$\alpha$/Fe]. We can, however, look into the star formation timescale of galaxies, i.e., the rate of the exponential decay of star formation, to understand how fast SF was quenched within the galaxy. In the bottom row of Fig. \ref{fig:MZR_matlasonly}, we colour-code our UDGs by their mass-weighted ages ($t_M$) and star formation timescales ($\tau$). We can see that UDGs that have the fastest star formation episodes (i.e., smallest $\tau$) are in the lowest regions of the plane, coinciding with the positioning of the older galaxies in the age panel. They are consistent with the evolving MZR at high-redshift, indicative of early-quenching, which is in agreement with the results from \citetalias{Ferre-Mateu_23}, and suggestive of a failed galaxy-like origin. On the other hand, UDGs that show prolonged SFHs (i.e., large $\tau$) are the ones that are younger and more metal-rich, aligned with a puffed-up dwarf formation scenario, as also found by \citetalias{Ferre-Mateu_23}.

\subsubsection{Dependence with GC richness and specific frequency}
\label{sec:MZR_GC}

\begin{figure}
    \centering
    \includegraphics[width=\columnwidth]{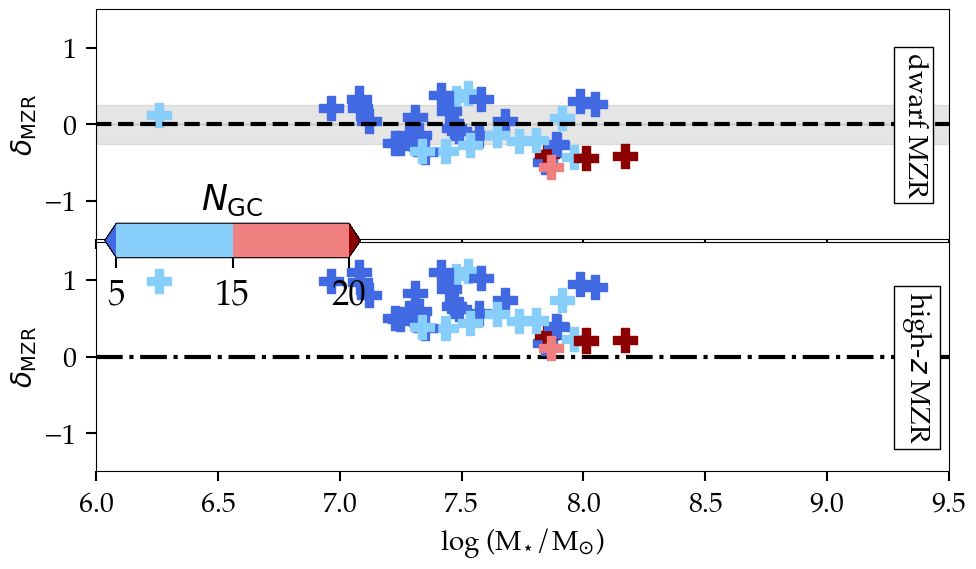} 
    \caption{Difference in metallicity ($\delta_{\rm MZR}$) between known MZRs and the MATLAS UDGs colour-coded by their number of GCs. Description is the same as Fig. \protect\ref{fig:simple_MZR}. GC--poor MATLAS UDGs follow the classical dwarf MZR from \protect\cite{Simon_19}. On another hand, all UDGs that follow the \protect\cite{Ma_15} MZR at high-redshift are GC-rich.}
    \label{fig:deltaMZR_NGC}
\end{figure}

One of the properties of UDGs that draws a lot of attention is the large number of GCs that some of them host in spite of their low stellar masses. Some extreme examples include DF44 (\citealt{vanDokkum_17}, but see also \citealt{Saifollahi_22} and \citealt{Forbes_Gannon_23}) and NGC 5846\_UDG1/MATLAS-2019 \citep[][]{Forbes_21,Mueller_20,Danieli_22}. 
Simulations and analytical models have tried to explain the existence of these many GCs in such low mass galaxies \citep[e.g., ][]{Carleton_21,Trujillo_Gomez_21,vanDokkum_22,Danieli_22,Doppel_23}.
In parallel, observational work has focused on understanding what are the main differences between GC--poor and GC--rich UDGs, in an attempt to correlate the higher fraction of GCs with other galaxy properties \citep{Forbes_20a, LaMarca_22}. 

In terms of stellar populations, \citetalias{Buzzo_22b} used SED fitting techniques to find evidence for an inherent difference between GC--rich and GC--poor UDGs. They found that GC--poor UDGs are consistent with the classical dwarf MZR from \cite{Simon_19}, having consistently higher metallicities than their GC--rich counterparts.
Interestingly, all of the UDGs that are consistent with the evolving MZR at $z=2.2$ proposed by \cite{Ma_15} are GC--rich, suggesting a formation scenario relying on early quenching for these galaxies.
Following up on this finding, \citetalias{Ferre-Mateu_23} studied the stellar populations of 25 UDGs, the largest spectroscopic sample to date, using data from Keck/KCWI.
They found that GC--poor UDGs consistently follow the classical dwarf MZR. However, they found that the GC--rich UDGs can be found both above and below the \cite{Simon_19} MZR. They suggested that the ones above it are likely tidal UDGs, while the ones below the MZR are consistent with early-quenching (failed galaxy scenarios), similar to what was suggested by \citetalias{Buzzo_22b}.

In the bottom row of Fig. \ref{fig:deltaMZR_NGC} we show the MATLAS UDGs colour-coded by their number of GCs. 
Differently from \citetalias{Buzzo_22b} and \citetalias{Ferre-Mateu_23}, we do not impose the hard separation that GC--rich UDGs are the ones with $N_{\rm GC} \geq 20$ and GC--poor UDGs the ones with fewer. Instead, the colour-scheme has been discretized so that evolutionary trends can be identified. 
In this figure, we see that GC--poor UDGs are overall more metal-rich galaxies, consistent with the classical dwarf MZR, and consistent with the recent findings of \cite{Jones_23} that gas-rich, field UDGs host few GCs. Alternatively, we see that all UDGs that follow the evolving MZR at high-redshift are GC--rich. We do not find the population of GC--rich metal-rich cluster UDGs found by \citetalias{Ferre-Mateu_23}, which may be expected as all of our UDGs are in groups/field, and thus less prone to tidal effects (see Fig. \ref{fig:MZR_matlasonly}). 

These results are in agreement with the recent findings of Pfeffer et al. (subm.), who used the EAGLE hydrodynamical simulations to find a correlation between the axis ratio and GC-richness of dwarf galaxies. In their study, round (higher $b/a$) dwarfs host more GCs than their elongated (low $b/a$) counterparts. As a a caveat, the works of \cite{Smith_13} and \cite{Smith_15} have shown that it is not possible to make a galaxy rounder through tidal stripping while also retaining a large number of GCs. If the harassment and stripping are strong enough to significantly alter their shapes, then the bulk of the GC system would be lost as well. In this study, however, we refer to galaxies that are round and GC--rich from their early stages.
As a sidenote, the effective radius of galaxies, not only UDGs, has been shown to correlate with the number of GCs by \cite{Harris_13}, also in agreement with our findings.

To understand if this trend between the metallicity and GC number is caused solely by the GC-richness or if it is being influenced by the mass of the galaxy, we also calculated the specific frequency per unit stellar mass of the galaxies, $T_N$ \citep{Zepf_93}. This specific frequency is defined as:

\begin{equation}
    T_N = \frac{N_{\rm GC}}{\left( \frac{M_{\star}}{10^9 M_{\odot}} \right)}.
\end{equation}

The MATLAS UDGs are colour-coded by their specific frequencies in the rightmost panel in the bottom row of Fig. \ref{fig:MZR_matlasonly}. We can see now that the trend between GC-richness and metallicity is not so clear anymore. Although on average metal-poor UDGs have higher specific frequencies, there is a non-negligible population of more metal-rich UDGs that also have high $T_N$ values. This indicates that the trend found between the metallicity and the GC number may be incorporating correlations with the stellar mass. It is important to note, nonetheless, that the UDGs that follow the high-redshift MZR all show high values of $T_N$.
This suggests that at least part of the reason that we do not find GC--rich, metal--rich UDGs is our lower stellar mass (on average) than \citetalias{Ferre-Mateu_23}. When controlling for the stellar mass of the galaxies using $T_N$, this population becomes apparent. These GC--rich metal--rich galaxies are likely tidal UDGs, as suggested by \citetalias{Ferre-Mateu_23}. This finding is reinforced by the fact that one confirmed tidal UDG in our sample, MATLAS-1824, has a similar metallicity and $T_N$ as these other GC--rich metal--rich UDGs. 

Further supporting these findings, \cite{Lim_18} have studied the specific frequencies ($S_N$) of UDGs in the Coma cluster and also found two different populations of UDGs in terms of their GC content. In their work, UDGs with high $S_N$ were found to be more dark matter dominated, with mass-to-light (M/L) ratios > 1000, while UDGs with low $S_N$ have M/L $\sim 500$, consistent with the predictions of failed galaxy and puffed-up dwarf scenarios, respectively.

Interestingly, a transitional population is also identified that has on average $\sim 10$ GCs ($10 < T_N < 200$); and these do not follow either the classical dwarf MZR nor the high-$z$ MZR (see light blue points in Fig. \ref{fig:deltaMZR_NGC}). 
Seeing a continuous trend in the GC numbers immediately leads to questioning whether GC--poor UDGs could evolve into GC--rich ones (or vice-versa). However, these galaxies, if we put together all of the properties studied so far, seem to be inherently different, not only in terms of their GC systems. 
GC--poor UDGs are primarily found in the field or low-density groups, they are young, are elongated, and show prolonged star formation histories.
GC--rich UDGs, on the other hand, are found in clusters or high-density groups, are primarily round, have old to ancient ages, and have quenched long ago, having formed the bulk of their stars in early and quick star formation episodes. 
Given their different properties, an evolutionary path between the two populations seems unlikely and, thus, two  (or more) formation scenarios seem to be necessary to explain these different types of UDGs \citep{Prole_19, Jones_23}. 

\subsection{Multiple classes of UDGs?}
\label{sec:two_types}

\begin{figure}
    \centering
    \includegraphics[width=\columnwidth,trim=1cm 1cm 2cm 0, clip]{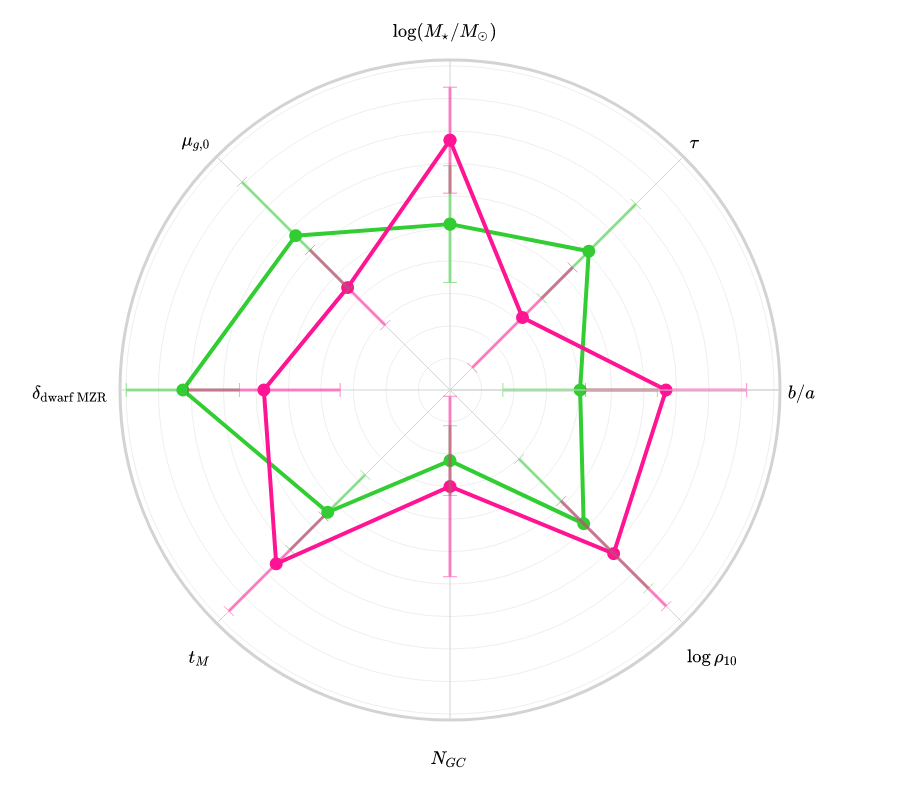}  
    \caption{The two groups of UDGs identified using the \texttt{KMeans} clustering algorithm.  In this figure, the radial axis shows the central value of a given property in each class, while the angular axis shows each property being considered by the clustering algorithm. This polar plot is to be read as the relative difference in the properties of UDGs belonging to the Classes A and B. Class A is shown in green and is comprised of UDGs that have lower stellar masses, prolonged SFHs (i.e., higher star formation timescales), are more elongated (lower $b/a$), live in less dense environments (lower $\log \rho_{10}$), host fewer GCs, are younger (smaller $t_M$), have higher $\delta_{\rm dwarf \, MZR}$ (i.e., closer to zero and consistent with the classical dwarf MZR), and are fainter (higher $\mu_{g,0}$). As these properties agree with the predictions of puffed-up dwarf-like formation scenarios, we associate Class A with UDGs having this origin. Class B is shown in pink and includes UDGs with higher stellar masses, rapid SFHs (i.e., smaller $\tau$), rounder (high $b/a$), living in the densest of our probed environments (higher $\log \rho_{10}$), hosting on average the biggest GC systems, older (higher $t_M$), with smaller $\delta_{\rm dwarf \, MZR}$ (i.e., negative values, being thus more metal-poor than the classical dwarf MZR), and brighter (smaller $\mu_{g,0}$). As these properties are aligned with the predictions of failed galaxy-like formation scenarios, we associate Class B with UDGs having formed through this pathway.}
    \label{fig:polar_two}
\end{figure}

\begin{figure*}
    \centering
    \includegraphics[width=\textwidth]{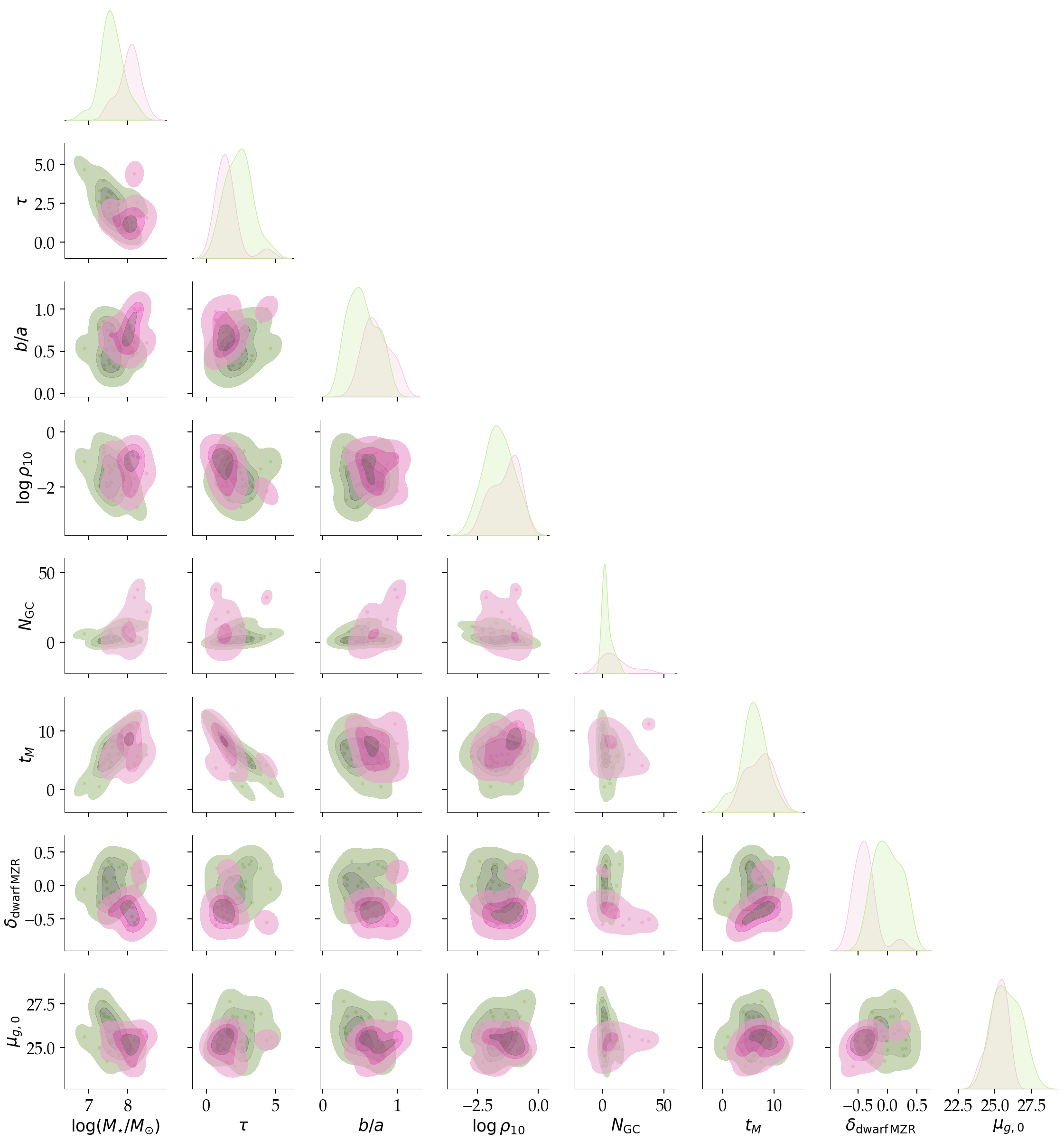}  
    \caption{The two groups of UDGs identified using the \texttt{KMeans} clustering algorithm. Distribution of UDGs in the parameter space defined by stellar mass ($\log (M_{\star}/M_{\odot})$), star formation timescale ($\tau$), axis ratio ($b/a$), local volume density ($\log \rho_{10}$), number of GCs ($N_{\rm GC}$), mass-weighted age ($t_M$), $\delta_{\rm dwarf \, MZR}$ (i.e., level of consistency between the UDGs and the classical dwarf MZR), and central surface brightness ($\mu_{g,0}$). Green points indicate galaxies that belong to class A (associated to the ``puffed-up dwarf'' scenario), pink points indicate galaxies that belong to class B (associated to the ``failed galaxy'' scenario). The histograms are smoothed using a Kernel Density Estimation (KDE) technique and indicate the distribution of a given parameter; green and pink histograms indicate those of class A and B, respectively.}
    \label{fig:two_types}
\end{figure*}

The results in the previous sections seemed to indicate that there exists at least two populations of UDGs. Driven by the idea that UDGs with different properties were formed by different processes, in what follows we investigate and compare the distribution and possible correlation between their global properties. Our goal is to identify classes of objects with similar properties and possibly associate them with one of the prominent formation scenarios for UDGs.

Assuming the existence of distinct populations of UDGs \citep{Forbes_20a}, and in order to better associate each galaxy to one of them, we applied the centroid-based clustering algorithm \texttt{KMeans} \citep{Macqueen_67} to the data. \texttt{KMeans} has been applied to Astronomy in various contexts and was proven to be a valuable methodology to classify large datasets \citep[see e.g.,][]{SanchezAlmeida_10,GarciaDias_18,Coccato_22}. The \texttt{KMeans} algorithm is an unsupervised learning method and is part of the Scikit-learn library \citep{scikit-learn}, responsible for implementing machine learning algorithms in Python. 

In our implementation, \texttt{KMeans} groups together galaxies that have similar properties in the multi-parameter space defined in our dataset and flags them accordingly into classes.
The initial parameters considered and compared here are stellar mass ($\log (M_{\star}/M_{\odot})$), star formation timescale ($\tau$), mass-weighted age ($t_M$), local volume density ($\log \rho_{10}$), number of GCs ($N_{\rm GC}$), GC specific frequency ($T_N$), central surface brightness ($\mu_{g,0}$), and the residual metallicity between the UDGs and the classical dwarf MZR ($\delta_{\rm dwarf \, MZR}$). The latter was included instead of metallicity in order to remove the degeneracy between metallicity and stellar mass so both properties could be analysed separately. Positive values of $\delta_{\rm dwarf \, MZR}$ mean that UDGs are above the dwarf MZR (i.e. more metal-rich), 0 is when they follow exactly the MZR, and negative values are for the cases where UDGs lie below the MZR (i.e., more metal-poor). The S\'ersic index ($n$) and effective radius ($R_{\rm e}$),  were not included in the analysis as they are degenerate with the surface brightness. Amongst our total sample of 59 galaxies, only 38 UDGs have all of these physical quantities available (i.e., 21 UDGs do not have GC counts). Therefore, these 38 are the ones used in our analysis. This is a small sample of UDGs and we caution that results obtained from it need to be further tested in a larger sample of galaxies, as these may not be representative of a diverse populations of UDGs and should not be extrapolated to such.
All properties were linearly scaled down to a range between 0 and 1 so that good measures of distance between classes could be assessed for every parameter analysed.
The \texttt{KMeans} clustering algorithm is isotropically applied to all parameters in the multiparameter space and, therefore, tends to produce round (rather than elongated) clusters. In this case, if unscaled data is provided (i.e., with unequal variances), then more weight is given to variables with smaller variance. Thus, rescaling the data is a necessary step to allow all properties to be equally examined by the algorithm.

We applied \texttt{KMeans} to the dataset allowing it to freely find the number of classes that best represents our data. We used the silhouette score quantity to measure the goodness of the clustering technique, where a high score means that the classes are well apart from each other and can be clearly distinguished. The score ranges from -1 to 1, where -1 means that sources are being associated with the wrong class, 0 means that classes are indistinguishable, and 1 means that the classes are perfectly separated. Two classes were found as the best number to describe our sample of UDGs, delivering the highest silhouette score (0.4) when compared to other numbers of classes.
\texttt{KMeans} allocates every single data point to the closest class centre in the multi-parameter space, which means that all UDGs analysed in this study are allocated to one of the classes. This does not mean that every UDG belonging to a particular class will have all of the properties compatible with that class, but that overall, most of their properties would be consistent. 
These classes were further assessed using Principal Component Analysis (PCA), which showed that two clusters are able to recover 60\% of the variance of the dataset. \texttt{DBScan}, another machine learning clustering algorithm, was also applied to the data and similarly returned that the best number of classes to represent this dataset is two, with a silhouette score of 0.3.

\begin{table}
    \centering
    \caption{Mean values of UDG classes obtained with the \texttt{KMeans} clustering results. The first block is the classification done using $N_{\rm GC}$ (only 38 galaxies). The second block is the classification without $N_{\rm GC}$ (all 59 MATLAS UDGs).}
    \begin{tabular}{ccc} \hline
        \multirow{2}{*}{\textbf{Parameter}} & \textbf{Class A} & \textbf{Class B} \\
         & \textit{(Puffed-up dwarf)} & \textit{(Failed galaxy)} \\ \hline
         Number & 23 & 15 \\
        $\log (M_{\star}/M_{\odot})$ & $7.31 \pm 0.17$ & $7.87 \pm 0.16$ \\
        $\tau$ (Gyr) & $2.27 \pm 0.75$ & $1.58 \pm 0.96$ \\
        $b/a$ & $0.59 \pm 0.19$ & $0.69 \pm 0.17$ \\
        $\log \rho_{10}$ (Mpc$^{-3}$) & $-1.53 \pm 0.56$ & $-1.36 \pm 0.59$ \\
        ${N}_{\rm GC}$ & $2.27 \pm 2.30$ & $13.09 \pm 11.01$\\ 
        $t_M$ (Gyr) &  $5.74 \pm 2.02$ & $7.34 \pm 2.36$ \\
        $\delta_{\rm dwarf \, MZR}$ (dex) & $-0.05 \pm 0.23$ & $-0.33 \pm 0.23$ \\
        $\mu_{g,0}$ (mag/arcsec$^2$) & $25.91 \pm 0.94$ & $25.19 \pm 0.62$\\ \hline \hline
        Number & 33 & 26 \\
        $\log (M_{\star}/M_{\odot})$ & $7.28 \pm 0.19$ & $7.71 \pm 0.27$ \\
        $\tau$ (Gyr) & $2.68 \pm 0.59$ & $1.54 \pm 0.81$ \\
        $b/a$ & $0.56 \pm 0.21$ & $0.67 \pm 0.16$ \\
        $\log \rho_{10}$ (Mpc$^{-3}$) & $-1.57 \pm 0.54$ & $-1.41 \pm 0.53$ \\
        % ${N}_{\rm GC}$ & $3.45 \pm 3.76$ & $10.57 \pm 11.44$\\ 
        $t_M$ (Gyr) &  $5.07 \pm 1.79$ & $7.25 \pm 2.20$ \\
        $\delta_{\rm dwarf \, MZR}$ (dex) & $-0.02 \pm 0.23$ & $-0.31 \pm 0.21$ \\
        $\mu_{g,0}$ (mag/arcsec$^2$) & $26.04 \pm 1.06$ & $25.33 \pm 0.64$\\ \hline
        % $R_{\rm e}$ (kpc) & $1.86 \pm 0.49$ & $1.84 \pm 0.64$ \\ \hline
    \end{tabular}
    \label{tab:two_types}
\end{table}

The two classes of UDGs are shown in Fig. \ref{fig:polar_two}. 
This plot shows the relative difference between the properties of the UDGs belonging to each class 
and provides an intuitive way of understanding which properties provide the biggest separation between classes, and properties that contribute less. 
From Fig. \ref{fig:polar_two}, we can see that the best discriminators of UDGs belonging to different classes are $\log (M_{\star}/M_{\odot})$, $\tau$, $t_M$, $\delta_{\rm dwarf \, MZR}$, and $\mu_{g,0}$. Properties that provide weaker separations are $b/a$, $\log \rho_{10}$, and $N_{\rm GC}$. 
Several iterations of this plot were analysed to remove properties that did not contribute to the separations, or that were degenerate with others properties already included and were, therefore, not adding anything new to the separation. Properties excluded were $R_{\rm e}$, $n$ and PA for showing no significant difference between groups. At this point, it is important to remind the reader that in machine learning algorithms, cleaning the dataset from any properties that are not contributing to the learning process is one of the most important tasks to recover reliable results. One might think that leaving parameters that return a null separation will not make a difference, but these add noise to the clustering process and could pollute and compromise the final recovered results. We emphasise, nonetheless, that the recovered classes are consistent independently of the addition of these parameters, however, the classes are found to have much larger uncertainties as the null parameters contaminate the clear separation provided by the significant ones. 

On the other hand, it is important to analyse carefully the properties that are providing the biggest separation to ensure these are physically meaningful and not just a result of selection effects or the properties being degenerate/correlated. For example, the surface brightness, effective radius, and number of GCs are known to correlate with the stellar mass. We thus carried out tests excluding every single one of these properties from the clustering algorithm to check whether the classes would change or not. We found that the recovered classes are similar regardless if any of these properties is excluded or included in the process, they however result in larger uncertainties on the class centres. 
Overall, in all iterations of applying \texttt{KMeans} to our dataset (i.e., with the inclusion/omission of properties) the classes maintained, except for one iteration. The only property capable of changing the identified classes was $T_N$, this is because even though all of the GC--rich galaxies have on average high specific frequencies, GC--poor UDGs show a mix of specific frequencies, capable of confusing the clustering algorithm and strongly changing the identified classes. Whether the separation with $T_N$ is more correct than the one without it needs to be further assessed in a larger sample of galaxies. In this study, we choose to use the results that are stable in the absence of $T_N$.
As a last comment, we note that if we exclude the GC number from the analysis, we can apply \texttt{KMeans} to the full sample of 59 UDGs (as the GC number is the only property that is not available for all galaxies). Once this is done, the exact same classes are identified, and the variance on the properties is smaller as the classes are comprised of a larger sample of galaxies. This gives us confidence that our clustering methodology is capturing the main properties of our dataset even when constrained to a smaller subsample.

The properties of the two classes of UDGs identified with \texttt{KMeans} as summarised in Table \ref{tab:two_types} are:
\begin{itemize}
    \item Class A. UDGs in this class have lower stellar masses, prolonged SFHs (i.e., higher star formation timescales), are more elongated (lower $b/a$), live in less dense environments (lower $\log \rho_{10}$), host fewer GCs, are younger (smaller $t_M$), have higher $\delta_{\rm dwarf \, MZR}$ (i.e., closer to zero and consistent with the classical dwarf MZR), and are fainter (higher $\mu_{g,0}$).

    \item Class B. UDGs in this class have higher stellar masses, rapid SFHs (i.e., smaller $\tau$), are rounder (high $b/a$), live in the densest of our probed environments (higher $\log \rho_{10}$), some host the biggest GC systems, are older (higher $t_M$), have smaller $\delta_{\rm dwarf \, MZR}$ (i.e., negative values, being thus more metal-poor than the classical dwarf MZR), and are brighter (smaller $\mu_{g,0}$).
\end{itemize}

The centres of the two identified classes and the number of UDGs associated with each class are shown in the upper block of Table \ref{tab:two_types} for the classification including $N_{GC}$. The lower block, on the other hand, shows the centres and number of UDGs following the classification applied to all 59 galaxies (i.e., removing $N_{\rm GC}$ from the analysis). The quoted uncertainties are the standard deviation of each analysed property considering the galaxies that belong to that class, i.e., an approximation of how far apart objects belonging to the same class can be. 
The cluster centres and uncertainties can also be seen in the histogram distribution of the properties in Fig. \ref{fig:two_types}. Properties that provide the best separation between classes will have more clear bimodal distributions, while the ones that are not strong discriminators will have similar Gaussian distributions. 
The classes each UDG in our sample are associated in the analyses with and without $N_{\rm GC}$ are given in Table \ref{tab:morphology}. It is reinforcing to see that 37 out of the 38 UDGs included in both analyses (i.e., with and without $N_{\rm GC}$) have preserved their classes. Only 1 UDG (MATLAS-984) changed classes when the number of GCs was excluded. We note that changing classes is expected if the classification is being driven by the property that was removed (i.e., $N_{\rm GC}$) and the other properties are in overlapping cluster regions. This is the case for MATLAS-984, which has properties that could fit into any class when $N_{\rm GC}$ is not considered (see Tables \ref{tab:morphology} and \ref{tab:stellarpops}).

We emphasise that these classes were obtained using a small sample of UDGs, all in low-density environments. Thus, the centres found may not be representative of a vast population of UDGs with strongly divergent properties. These cluster centres should perhaps be interpreted as a measurement of which properties are important discriminators of UDG formation scenarios and which ones are not. 
Although there is not a net separation or threshold in the properties of UDGs formed via the two scenarios, and possibly not a one-to-one correspondence between a given class and a formation scenario, we recognise that UDGs in Class A have properties (see Table \ref{tab:two_types}) that are more aligned with being formed through puffed-up dwarf scenarios, i.e., properties similar to those of classical dwarf galaxies. On the other hand, UDGs in Class B share similar properties to those of failed galaxy scenarios (see Table \ref{tab:two_types}), i.e., properties consistent with early-quenching. We note that tidal UDGs, due to low number statistics, was not found as a third class in our dataset, although these types of UDGs have overall different properties than puffed-up dwarfs or failed galaxies. This class is expected to become clearer in larger samples of UDGs, including more massive and cluster ones, where tidal effects are more present. 

This clustering algorithm was applied to a small set of galaxies and needs to be further tested, both with a larger number of UDGs and in comparison with classical dwarfs. This way, a more comprehensive comparison between subpopulations can be made. We note, nonetheless, that this separation (in terms of GC number, axis ratio, and mass) has also been reported in the Coma cluster \citep{Lim_18}, so these two classes of UDGs may be a general separation regardless of their environments. 
A study combining and comparing the MATLAS UDGs, the UDGs studied in \citetalias{Buzzo_22b} and a control sample of classical MATLAS dwarfs is under preparation (Buzzo et al. in prep.) and will focus on deepening these clustering findings. Other properties can also be added to the clustering in the future to enrich the classification, namely [$\alpha$/Fe] \citep[e.g., ][\citetalias{Ferre-Mateu_23}]{Ferre-Mateu_18,Villaume_22}, dynamical masses and velocity dispersions \citep[e.g., ][]{Gannon_20,Gannon_21,Gannon_22,Gannon_23}, positioning in the phase-space diagram \citep[e.g., ][\citetalias{Ferre-Mateu_23}]{Gannon_22,Forbes_23}, etc.
As a final remark, once the classes are further tested and if they are proven to be well founded, they could be used to predict unknown properties of UDGs. For example, by knowing the stellar populations and morphology of a given UDG, using these clustering results, one would be able to predict if it is GC--poor or GC--rich, in cases where knowing the GC--richness of the galaxy would otherwise be impossible (e.g., at distance higher $>$ 100 Mpc), further showing the importance of such findings.

\section{Conclusions}
\label{sec:conclusions}

In the current study, we have used the fully Bayesian Monte Carlo Markov Chain inference code \texttt{Prospector} to perform spectral energy distribution fitting on the 59 MATLAS UDGs using data from the optical to the mid-IR.
We use \texttt{Prospector} to recover stellar populations using two different configurations, one with the redshift as a free parameter, and one with the redshift fixed to the redshift of the group where the UDG was identified.

We find that the models with the redshift fixed to that of the host group are the most appropriate to study these galaxies. This assumption is based on the results of \cite{Heesters_23}, who found that 75\% of the MATLAS dwarfs studied by them are at the same distance as their massive hosts. This is supported by our study with Keck/DEIMOS of three MATLAS UDGs, also found to be at the same redshift as their hosts. 

Our results indicate that the MATLAS UDGs have intermediate-to-old ages, with an average mass-weighted age of $7.1 \pm 1.8$ Gyr. They are also metal-poor with an average [M/H] of $-1.2 \pm 0.2$ dex. They have an average star formation timescale $\tau$ of $1.6 \pm 0.7$ Gyr and are consistent with no dust attenuation, displaying an average of $A_{V} = 0.12 \pm 0.07$ mag. 

When studying the MATLAS UDGs in the mass--metallicity plane, we find that their distribution is highly bimodal, with the first mode being well explained by the classical dwarf MZR, while the second one is more consistent with the evolving MZR at a high-redshift. 

To further investigate the cause of this bimodality, we study the positioning of the UDGs in the mass--metallicity plane according to their surface brightness, effective radii, axis ratios, local volume densities, mass-weighted ages, star formation timescales, number of GCs and GC specific frequencies. We find that UDGs split into two main classes: 

\textbf{Class A:} They have lower stellar masses, prolonged SFHs, are more elongated, live in less dense environments, host fewer GCs, are younger, are consistent with the classical dwarf MZR, and are fainter. 

\textbf{Class B:} UDGs in this class have higher stellar masses, rapid SFHs, are rounder, live in the densest of our probed environments, host on average the biggest GC systems, are older, lie below the classical dwarf MZR, and are brighter. These galaxies are overall better explained by the evolving MZR at high-redshift, i.e., consistent with early-quenching.

This overall picture seems to indicate that UDGs belonging to Class A, combining all of the aforementioned properties, are better explained by puffed-up dwarf formation scenarios. Alternatively, UDGs of Class B, taking into consideration all of the morphological and stellar population properties found, seem to be more aligned with failed galaxy formation scenarios, in agreement with the spectroscopic studies. Further testing of this clustering technique, including a larger sample of UDGs and classical dwarfs for comparison is under preparation (Buzzo et al. in prep.).

This paper provides a continuation of the photometric study of the stellar populations of UDGs across the sky. We demonstrate that SED fitting techniques, coupled with a broad wavelength coverage, are an important approach to statistically understand the origins of UDGs and their impact on the field of galaxy formation and evolution.

\section*{Acknowledgements}
We thank the anonymous referee for suggestions that greatly improved the manuscript.

We are thankful to Patrick Durrell for his contributions to the MATLAS survey. We thank Pavan Uttarkar for all the help with \texttt{KMeans} and other clustering algorithms tested and used in this work. We also thank Amelia Fraser-McKelvie for her colloquium at Swinburne University which gave us the idea of applying clustering algorithms to our dataset. We thank Yimeng Tang for important discussions about Prospector's configurations and outputs.

This research was supported by the Australian
Research Council Centre of Excellence for All Sky Astrophysics in 3 Dimensions (ASTRO 3D), through project number CE170100013. DAF, JPB, JP and WJC thank the ARC for support via DP220101863. AJR was supported by National Science Foundation grant AST-2308390. AFM has received support from CEX2019-000920-S, RYC2021-031099-I and PID2021-123313NA-I00 of MICIN/AEI/10.13039/501100011033/FEDER,UE, NextGenerationEU/PRT.
S.L. acknowledges the support from the Sejong Science Fellowship Program through the National Research Foundation of Korea (NRF-2021R1C1C2006790).

This paper is based in part on observations from the Legacy Survey, which consists of three individual and complementary projects: the Dark Energy Camera Legacy Survey (DECaLS; Proposal ID \#2014B-0404; PIs: David Schlegel and Arjun Dey), the Beijing-Arizona Sky Survey (BASS; NOAO Prop. ID \#2015A-0801; PIs: Zhou Xu and Xiaohui Fan), and the Mayall z-band Legacy Survey (MzLS; Prop. ID \#2016A-0453; PI: Arjun Dey).
This publication makes use of data products from the Wide-field Infrared Survey Explorer, which is a joint project of the University of California, Los Angeles, and the Jet Propulsion Laboratory/California Institute of Technology, funded by the National Aeronautics and Space Administration.

\textit{Software:} astropy \citep{Astropy_13,Astropy_18}, FSPS \citep{Conroy_10a, Conroy_10b}, python-fsps \citep{Johnson_21a}, \texttt{Prospector} \citep{Leja_17,Johnson_21}.

%%%%%%%%%%%%%%%%%%%%%%%%%%%%%%%%%%%%%%%%%%%%%%%%%%
\section*{Data Availability}
DECaLS data are available via the  \href{https://www.legacysurvey.org/decamls/}{Legacy survey portal}. \textit{WISE} data are available via the \href{https://wise2.ipac.caltech.edu/docs/release/allsky/}{WISE archive}. MATLAS data are available from the CDS archives and through the \href{http://matlas.astro.unistra.fr}{MATLAS portal}.



\bibliographystyle{mnras}
\bibliography{bibli} % if your bibtex file is called example.bib

%\begin{thebibliography}{99}
%\end{thebibliography}

%%%%%%%%%%%%%%%%%%%%%%%%%%%%%%%%%%%%%%%%%%%%%%%%%%

%%%%%%%%%%%%%%%%% APPENDICES %%%%%%%%%%%%%%%%%%%%%

\appendix

\section{DEIMOS spectroscopic data}
\label{sec:appendix_deimos}

\begin{table}
\centering
\caption{\texttt{pPXF} fitting results for the MATLAS UDGs studied in this work. Columns stand for (1) Galaxy ID; (2) Signal-to-noise ratio of the resulting spectrum; (3) Recovered recessional velocity; (4) Name of Host;  (5) Recessional velocity of host galaxy.}
\scalebox{0.85}{
\begin{tabular}{ccccccc} \hline
\textbf{Galaxy} & \textbf{S/N (\AA$^{-1}$)} & \textbf{V (km s$^{-1}$)} & \textbf{Candidate Host} & \textbf{V$_{\rm host}$ (km s$^{-1}$)} \\ \hline
MATLAS-342 & 8.2  & $2142 \pm 31$ & NGC 2481 & 2153 \\ 
MATLAS-368 & 3.8  & $2035 \pm 97$ & NGC 2577 & 2073 \\
MATLAS-1059 & 3.2  & $1248 \pm 360$ & NGC 3683 & 1700 \\ \hline
\end{tabular}}
\label{tab:ppxf_matlas}
\end{table}

Our spectroscopic data were obtained from the Keck II telescopes with the Deep Imaging Multi-Object Spectrograph \citep[DEIMOS,][]{Faber_03} spectrograph.
MATLAS-342, MATLAS-368, and MATLAS-1059 were the selected targets due to their higher surface brightness and/or presence of a bright nucleus. They were observed with DEIMOS on the nights of 2022 November 23 and 2022 November 24, with good seeing ranging from 0.4'' -- 0.8'' most of the nights, with a short interval of seeing $\sim$ 2''. MATLAS-342 and MATLAS-368 were observed with the blue-sensitive 1200 lines mm$^{-1}$ grating (1200B), GG400 filter, central wavelength 5850 {\AA} and wavelength coverage of 4500 - 7200 {\AA} for one hour, with individual exposures of 1800 seconds with a custom-designed 3 arcsec-wide longslit. MATLAS-1059 was observed with the same setup, but only for 1800 seconds.

The data were reduced with the open-source reduction package \texttt{PypeIt} \citep[version 1.12.0,][]{Prochaska_20}. The arc image obtained in the same night as our observations turned out to have signs of ghosting, and thus, additional arc images were obtained in the same configurations on the night 2023 March 23 and were the ones used to reduce this data.

\begin{figure}
    \centering
    \includegraphics[width=\columnwidth]{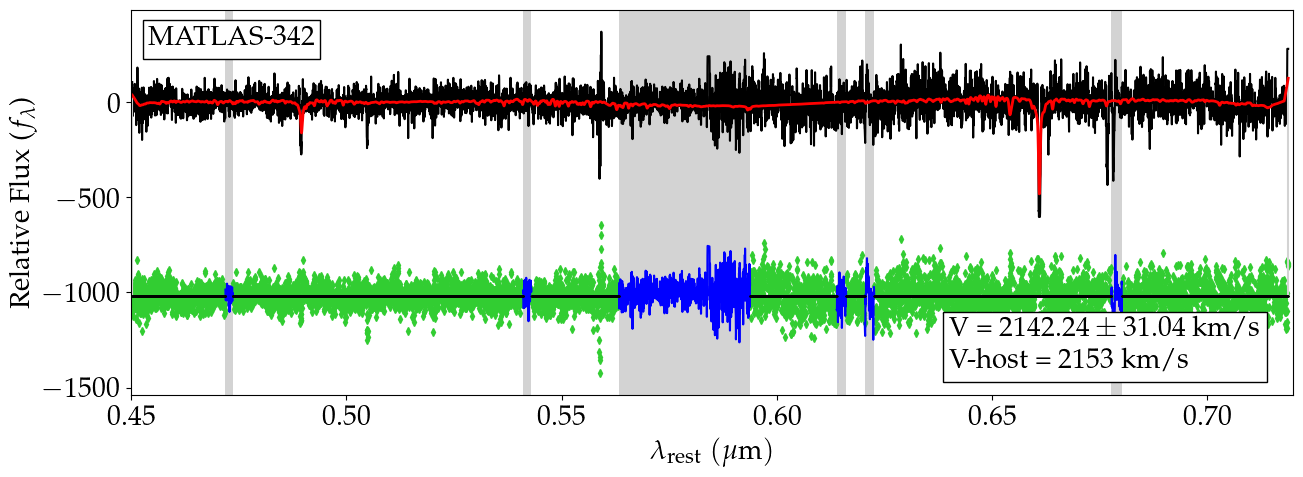}
    \caption{\texttt{pPXF} fitting of MATLAS-342, one MATLAS UDG with available spectra from Keck/DEIMOS. Black curve shows the spectrum of each source, while red curve shows the best-fit. Green dots are the residuals from the fit. Grey regions show the masked areas in the fit. Legend box within the plot shows the recovered radial velocity of the UDG ($V = 2142 \pm 31$ km/s) and the radial velocity of its host galaxy ($V = 2153$ km/s). Results indicate that this UDG is at the same redshift as its host.}
    \label{fig:ppxf_matlas}
\end{figure}

We fit each spectrum with pPXF \citep{Cappellari_17}, using the MILES stellar library \citep{Vazdekis_15}, assuming a Kroupa initial mass function \citep{Kroupa_01} and the BaSTI isochrones. 
% We use a single fit with an additional 4th-order additive and 5th-order multiplicative polynomial. 
In Fig. \ref{fig:ppxf_matlas}, we show one example of \texttt{pPXF} fit for the galaxy MATLAS-342. Results from the fits of all three galaxies are summarised in Table \ref{tab:ppxf_matlas}. The results indicate that all three probed UDGs are at the same redshifts as their massive neighbours.

We note that MATLAS-368 was selected to be followed up because it was thought to be nucleated, which would facilitate recovering its velocity because of its brightness. However, the central point source was found to be at $z=0$, thus likely being a foreground star. As an alternative, we studied the galaxy light coming from the outer parts of the slit, but this rendered low S/N data. The identification of the H$\alpha$ line was possible, resulting in the recessional velocity shown in Table \ref{tab:ppxf_matlas}. We caution, nonetheless, that the result is based on a single line, which is highly uncertain.

The spectrum of the galaxy that was found to be closer than its host, MATLAS-1059, has very low signal-to-noise (3.8 \AA$^{-1}$) due to its low exposure time, resulting in a much more uncertain velocity than the other studied UDGs. Even in this extreme case of low S/N, the recovered velocity is consistent with the one of the host. Additionally, from CFHT and DECaLS imaging, we can see that this UDG seems tidally disrupted by the host, further confirming its proximity to it.

These findings show that all three UDGs studied are at the redshifts of their hosts. This is in agreement with the recent study of \cite{Heesters_23} who found, after studying 56 MATLAS dwarfs, that 75\% of them were indeed at the same redshifts as their hosts, including the three UDGs present in their sample. Putting together our results and the ones from \cite{Heesters_23}, we have a cumulative result that 6 out of 6 MATLAS UDGs followed up with spectroscopy are confirmed as associated with their parent host galaxy. We thus conclude that assuming the redshift of the host for the UDGs that have not been spectroscopically followed-up is a reasonable assumption, bearing in mind that these are based on ensemble properties and may not hold for every individual galaxy. 

\section{Tables}

\subsection{MATLAS UDGs Photometry}
\label{sec:photometry}

\renewcommand{\arraystretch}{1.3}
\begin{table*}
\scalebox{0.77}{
\begin{threeparttable}
\caption{Optical, near- and mid-IR photometry of the MATLAS UDGs.}
\begin{tabular}{cccccccccccccccccccc} \hline
\multirow{2}{*}{ID} & $g$ & $r$ & $g-r$ & $i$ & $g-i$ & $z$ & $g-z$ & $W1$ & $W2$ & $W3$ & $W4$ \\ 
& [mag] & [mag] & [mag] & [mag] &  [mag] & [mag] & [mag] & [mag] & [mag] & [mag] & [mag] \\ \hline
MATLAS-42 & $17.46\pm0.02$ & $17.03\pm0.03$ & 0.43 & -- & -- & $16.70\pm0.03$ & 0.76 & $17.10\pm0.15$ & $17.92\pm0.24$ & $>18.01$ & -- \\
MATLAS-48 & $18.94\pm0.05$ & $18.26\pm0.07$ & 0.67 & -- & -- & $18.00\pm0.07$ & 0.94 & $18.39\pm0.17$ & $18.92\pm0.13$ & -- & $>18.57$ \\
MATLAS-141 & $20.09\pm0.07$ & $19.60\pm0.08$ & 0.50 & -- & -- & $19.33\pm0.12$ & 0.76 & $19.24\pm0.26$ & $20.97\pm0.77$ & $>20.86$ & $>17.29$ \\
MATLAS-149 & $20.10\pm0.07$ & $19.66\pm0.08$ & 0.44 & -- & -- & $19.41\pm0.16$ & 0.68 & $19.91\pm0.13$ & $20.98\pm1.52$ & $>18.21$ & $>16.41$ \\
MATLAS-177 & $19.18\pm0.08$ & $18.52\pm0.06$ & 0.66 & -- & -- & $>19.73$ & -- & $18.76\pm0.06$ & $18.97\pm0.21$ & $>20.30$ & $>16.18$ \\
MATLAS-262 & $19.96\pm0.09$ & $19.27\pm0.08$ & 0.69 & $19.03\pm0.32$ & 0.93 & $18.95\pm0.16$ & 1.01 & $19.86\pm0.68$ & $20.18\pm1.38$ & -- & $>15.81$ \\
MATLAS-342 & $18.78\pm0.03$ & $18.18\pm0.03$ & 0.60 & $17.79\pm0.04$ & 0.99 & $17.37\pm0.05$ & 1.41 & $18.54\pm0.18$ & $18.82\pm0.80$ & -- & $>19.32$ \\
MATLAS-365 & $19.75\pm0.09$ & $19.15\pm0.07$ & 0.59 & $18.96\pm0.08$ & 0.79 & $19.12\pm0.16$ & 0.63 & $19.18\pm0.32$ & $19.86\pm1.47$ & $>17.79$ & $>17.37$ \\
MATLAS-368 & $18.06\pm0.08$ & $17.50\pm0.08$ & 0.55 & $17.39\pm0.08$ & 0.67 & $17.29\pm0.13$ & 0.76 & $18.56\pm0.08$ & $19.58\pm1.08$ & $>19.23$ & $>15.03$ \\
MATLAS-405 & $18.88\pm0.05$ & $18.31\pm0.05$ & 0.57 & -- & -- & $>19.76$ & -- & $18.43\pm0.20$ & $19.11\pm0.81$ & $>21.73$ & $>19.95$ \\
MATLAS-478 & $17.69\pm0.02$ & $16.91\pm0.02$ & 0.78 & -- & -- & $>17.60$ & -- & $16.74\pm0.09$ & $16.86\pm0.04$ & -- & $>15.42$ \\
MATLAS-585 & $18.72\pm0.03$ & $18.19\pm0.03$ & 0.53 & $17.90\pm0.04$ & 0.82 & $17.87\pm0.08$ & 0.85 & $18.46\pm0.18$ & $18.94\pm0.21$ & $>20.83$ & -- \\
MATLAS-658 & $20.53\pm0.25$ & $19.77\pm0.28$ & 0.76 & $19.20\pm0.50$ & 1.32 & $19.10\pm0.50$ & 1.43 & $19.09\pm0.21$ & $19.56\pm0.41$ & -- & -- \\
MATLAS-799 & $17.40\pm0.03$ & $16.73\pm0.03$ & 0.67 & $16.51\pm0.02$ & 0.88 & $16.48\pm0.04$ & 0.92 & $16.85\pm0.02$ & $17.44\pm0.06$ & -- & -- \\
MATLAS-898 & $18.15\pm0.05$ & $17.58\pm0.05$ & 0.57 & $17.43\pm0.05$ & 0.72 & $17.21\pm0.07$ & 0.94 & $17.94\pm0.06$ & $18.22\pm0.80$ & $>20.68$ & -- \\
MATLAS-951 & $17.97\pm0.06$ & $17.18\pm0.05$ & 0.79 & $16.87\pm0.05$ & 1.10 & $16.54\pm0.06$ & 1.43 & $17.78\pm0.26$ & $18.09\pm0.10$ & -- & -- \\
MATLAS-984 & $19.60\pm0.04$ & $18.93\pm0.05$ & 0.67 & -- & -- & $>19.75$ & -- & $18.69\pm0.03$ & $20.53\pm0.25$ & $>18.52$ & $>16.76$ \\
MATLAS-1059 & $17.88\pm0.05$ & $17.17\pm0.05$ & 0.72 & -- & -- & $>18.34$ & -- & $17.69\pm0.10$ & $18.35\pm0.31$ & $>23.98$ & $>15.99$ \\
MATLAS-1174 & $19.47\pm0.16$ & $18.96\pm0.13$ & 0.51 & $18.64\pm0.10$ & 0.83 & $18.40\pm0.18$ & 1.07 & $19.09\pm0.19$ & $19.30\pm1.14$ & -- & $>19.31$ \\
MATLAS-1177 & $18.29\pm0.01$ & $17.54\pm0.01$ & 0.74 & -- & -- & $>19.51$ & -- & $17.76\pm0.03$ & $17.91\pm0.09$ & $>16.63$ & $>15.62$ \\
MATLAS-1205 & $19.49\pm0.09$ & $18.94\pm0.09$ & 0.55 & $18.60\pm0.08$ & 0.89 & $18.71\pm0.14$ & 0.78 & $19.85\pm0.25$ & $20.81\pm0.80$ & $>18.59$ & -- \\
MATLAS-1216 & $19.05\pm0.27$ & $18.96\pm0.16$ & 0.08 & $18.71\pm0.28$ & 0.34 & $18.33\pm0.19$ & 0.72 & $19.19\pm0.61$ & $20.07\pm1.27$ & $>20.23$ & $>18.39$ \\
MATLAS-1225 & $18.51\pm0.11$ & $17.74\pm0.12$ & 0.78 & $17.17\pm0.11$ & 1.35 & $16.99\pm0.17$ & 1.53 & $17.37\pm0.05$ & $17.94\pm0.20$ & $>20.83$ & -- \\
MATLAS-1245 & $19.58\pm0.05$ & $18.93\pm0.05$ & 0.64 & $18.80\pm0.11$ & 0.78 & $18.69\pm0.15$ & 0.88 & $18.97\pm0.43$ & -- & $>20.62$ & $>17.98$ \\
MATLAS-1246 & $17.88\pm0.05$ & $17.23\pm0.03$ & 0.64 & $17.04\pm0.03$ & 0.84 & $16.94\pm0.07$ & 0.94 & $17.47\pm0.09$ & $18.73\pm0.65$ & $>17.42$ & $>16.21$ \\
MATLAS-1248 & $17.49\pm0.01$ & $16.86\pm0.01$ & 0.63 & $16.64\pm0.01$ & 0.85 & $16.53\pm0.03$ & 0.96 & $16.89\pm0.09$ & $17.50\pm0.06$ & $>16.73$ & $>18.37$ \\
MATLAS-1249 & $19.18\pm0.04$ & $19.00\pm0.04$ & 0.17 & $18.82\pm0.05$ & 0.35 & $18.53\pm0.07$ & 0.65 & $18.02\pm0.03$ & $18.31\pm0.23$ & -- & $>15.97$ \\
MATLAS-1262 & $18.71\pm0.02$ & $18.11\pm0.02$ & 0.60 & $17.88\pm0.03$ & 0.83 & $17.82\pm0.07$ & 0.89 & $18.33\pm0.17$ & $19.49\pm1.03$ & -- & $>16.20$ \\
MATLAS-1274 & $17.52\pm0.02$ & $17.03\pm0.02$ & 0.50 & $16.86\pm0.02$ & 0.66 & $16.68\pm0.06$ & 0.84 & $17.55\pm0.09$ & $18.36\pm0.48$ & -- & $>19.61$ \\
MATLAS-1302 & $20.73\pm0.07$ & $20.13\pm0.07$ & 0.59 & $19.82\pm0.06$ & 0.91 & $19.63\pm0.14$ & 1.10 & $20.10\pm0.24$ & $20.33\pm0.82$ & $>18.46$ & $>16.50$ \\
MATLAS-1337 & $18.47\pm0.04$ & $18.22\pm0.04$ & 0.25 & $18.08\pm0.04$ & 0.39 & $17.82\pm0.09$ & 0.65 & $17.74\pm0.17$ & $18.04\pm0.37$ & $>16.84$ & $>14.61$ \\
MATLAS-1413 & $18.43\pm0.13$ & $18.11\pm0.15$ & 0.32 & $17.84\pm0.21$ & 0.59 & $17.65\pm0.28$ & 0.78 & $18.55\pm0.35$ & $18.71\pm0.43$ & -- & -- \\
MATLAS-1493 & $19.71\pm0.08$ & $19.28\pm0.08$ & 0.43 & $19.04\pm0.08$ & 0.67 & $18.96\pm0.14$ & 0.75 & $20.43\pm1.92$ & $20.86\pm1.64$ & -- & -- \\
MATLAS-1494 & $18.81\pm0.05$ & $18.29\pm0.01$ & 0.51 & $18.04\pm0.03$ & 0.77 & $17.94\pm0.08$ & 0.87 & $19.18\pm1.44$ & $20.03\pm1.58$ & -- & $>19.67$ \\
MATLAS-1534 & $20.26\pm0.10$ & $19.58\pm0.08$ & 0.67 & -- & -- & $>20.45$ & -- & $19.11\pm0.60$ & $19.85\pm1.00$ & $>19.02$ & $>16.39$ \\
MATLAS-1550 & $19.71\pm0.05$ & $19.07\pm0.06$ & 0.64 & -- & -- & $>20.38$ & -- & $19.66\pm0.32$ & -- & -- & $>17.28$ \\
MATLAS-1558 & $19.59\pm0.06$ & $19.12\pm0.06$ & 0.47 & -- & -- & $>20.05$ & -- & $19.92\pm0.14$ & $20.15\pm0.50$ & $>23.39$ & -- \\
MATLAS-1589 & $19.98\pm0.13$ & $19.59\pm0.14$ & 0.40 & -- & -- & $19.05\pm0.93$ & 0.93 & $19.85\pm0.39$ & $20.05\pm0.30$ & $>18.85$ & -- \\
MATLAS-1606 & $18.36\pm0.05$ & $17.69\pm0.05$ & 0.67 & -- & -- & $>19.74$ & -- & $18.15\pm0.18$ & $19.01\pm0.66$ & $>18.83$ & $>23.44$ \\
MATLAS-1615 & $18.26\pm0.04$ & $17.82\pm0.03$ & 0.44 & -- & -- & $>19.21$ & -- & $18.16\pm0.30$ & $18.78\pm0.61$ & $>21.47$ & -- \\
MATLAS-1616 & $18.65\pm0.06$ & $18.08\pm0.05$ & 0.57 & -- & -- & $>19.50$ & -- & $18.92\pm0.18$ & $19.48\pm0.71$ & -- & $>15.50$ \\
MATLAS-1630 & $19.28\pm0.08$ & $18.23\pm0.07$ & 1.05 & -- & -- & $>19.74$ & -- & $18.61\pm0.17$ & $19.36\pm0.15$ & $>17.27$ & $>16.97$ \\
MATLAS-1647 & $20.80\pm0.06$ & $20.17\pm0.07$ & 0.63 & -- & -- & $>20.31$ & -- & $20.37\pm0.15$ & $20.21\pm0.93$ & -- & -- \\
MATLAS-1779 & $18.66\pm0.09$ & $18.03\pm0.08$ & 0.63 & $18.00\pm0.12$ & 0.66 & $17.70\pm0.17$ & 0.96 & $18.67\pm0.68$ & $19.60\pm1.56$ & $>22.42$ & $>20.48$ \\
MATLAS-1794 & $19.02\pm0.02$ & $18.57\pm0.01$ & 0.44 & $18.38\pm0.04$ & 0.63 & $18.24\pm0.02$ & 0.78 & $19.01\pm0.30$ & -- & $>21.14$ & -- \\
MATLAS-1824 & $18.36\pm0.04$ & $17.87\pm0.05$ & 0.50 & -- & -- & $>19.42$ & -- & $18.22\pm0.36$ & $19.27\pm0.32$ & $>18.60$ & -- \\
MATLAS-1847 & $19.28\pm0.06$ & $18.66\pm0.05$ & 0.62 & $18.36\pm0.07$ & 0.91 & $18.82\pm0.11$ & 0.46 & $19.65\pm0.30$ & -- & $>16.72$ & -- \\
MATLAS-1855 & $18.97\pm0.05$ & $18.44\pm0.05$ & 0.54 & $18.34\pm0.06$ & 0.63 & $18.30\pm0.10$ & 0.67 & $19.77\pm1.66$ & -- & -- & $>15.90$ \\
MATLAS-1865 & $18.93\pm0.02$ & $18.39\pm0.02$ & 0.54 & -- & -- & $>18.88$ & -- & $18.83\pm0.12$ & $19.60\pm0.64$ & -- & $>19.08$ \\
MATLAS-1907 & $18.11\pm0.01$ & $17.57\pm0.01$ & 0.53 & $17.30\pm0.02$ & 0.80 & $17.20\pm0.02$ & 0.91 & $17.74\pm0.13$ & $18.07\pm0.34$ & $>22.37$ & $>20.08$ \\
MATLAS-1957 & $18.17\pm0.04$ & $17.43\pm0.04$ & 0.74 & $17.25\pm0.07$ & 0.91 & $17.07\pm0.08$ & 1.10 & $17.01\pm0.04$ & $17.75\pm0.03$ & -- & $>19.91$ \\
MATLAS-1975 & $19.08\pm0.11$ & $18.50\pm0.10$ & 0.59 & $18.06\pm0.17$ & 1.02 & $17.80\pm0.16$ & 1.29 & $19.14\pm0.55$ & $19.14\pm0.94$ & $>21.64$ & $>20.86$ \\
MATLAS-1985 & $18.06\pm0.10$ & $17.26\pm0.07$ & 0.80 & $17.03\pm0.19$ & 1.03 & $16.85\pm0.20$ & 1.21 & $18.04\pm0.28$ & -- & -- & $>16.60$ \\
MATLAS-1991 & $17.92\pm0.02$ & $17.36\pm0.01$ & 0.56 & $17.08\pm0.03$ & 0.84 & $16.98\pm0.03$ & 0.95 & $17.56\pm0.09$ & $18.35\pm2.00$ & $>18.82$ & $>18.24$ \\
MATLAS-1996 & $19.91\pm0.11$ & $19.47\pm0.10$ & 0.44 & $19.20\pm0.21$ & 0.70 & $19.08\pm0.22$ & 0.82 & $22.35\pm10.7$3 & $23.41\pm2.51$ & $>18.28$ & -- \\
MATLAS-2019 & $17.82\pm0.02$ & $17.20\pm0.02$ & 0.61 & $16.95\pm0.04$ & 0.87 & $16.81\pm0.04$ & 1.01 & $18.13\pm0.14$ & $18.45\pm0.19$ & -- & $>16.54$ \\
MATLAS-2021 & $19.37\pm0.11$ & $18.79\pm0.09$ & 0.58 & $18.74\pm0.25$ & 0.63 & $18.49\pm0.21$ & 0.88 & $24.23\pm1.86$ & -- & -- & $>18.53$ \\
MATLAS-2103 & $17.94\pm0.01$ & $17.41\pm0.01$ & 0.53 & $17.16\pm0.04$ & 0.78 & $17.08\pm0.03$ & 0.86 & $18.06\pm0.13$ & $18.41\pm1.95$ & $>19.70$ & $>20.36$ \\
MATLAS-2184 & $19.59\pm0.08$ & $19.07\pm0.10$ & 0.52 & -- & -- & $>19.30$ & -- & $21.62\pm1.17$ & -- & $>22.56$ & $>15.93$ \\
\hline
\end{tabular}
\begin{tablenotes}
      \small
      \item \textbf{Note.} Columns are: (1) Galaxy ID; (2) \texttt{GALFITM} DECaLS $g$-band magnitude; (3) \texttt{GALFITM} DECaLS $r$-band magnitude; (4) $g-r$ colour; (5) \texttt{GALFITM} DECaLS $i$-band magnitude; (6) $g-i$ colour; (7) \texttt{GALFITM} DECaLS $z$-band magnitude; (8) $g-z$ colour; (9) \textit{WISE} $3.4 \mu$-band magnitude; (10) \textit{WISE} $4.6 \mu$-band magnitude; (11) \textit{WISE} $12 \mu$-band magnitude; (12) \textit{WISE} $22 \mu$-band magnitude. '--' stands for unavailable data. '>' denote upper limit magnitudes.
\end{tablenotes}
\label{tab:photometry}
\end{threeparttable}}
\end{table*}

\subsection{\texttt{GALFITM} results and physical properties of MATLAS UDGs}
\label{sec:appendix_galfitm}

\renewcommand{\arraystretch}{1.3}
\begin{table*}
\scalebox{0.78}{
\begin{threeparttable}
\caption{\texttt{GALFITM} morphological parameters and physical properties of the MATLAS UDGs.}
\begin{tabular}{cccccccccccccccccc} \hline
\multirow{2}{*}{ID} & $R_{\rm e}$ & $n$ & $b/a$ & PA & $R_{\rm e}$ & $\langle \mu_{g,\textrm{e}} \rangle$ & $\mu_{g,0}$ & Candidate Host & Host Distance & \texttt{KMeans} Class \\ 
& [arcsec] & & & [degrees] & [kpc] & [mag/arcsec$^2$] & [mag/arcsec$^2$] & & [Mpc] & [with $N_{\rm GC}$, w/o $N_{\rm GC}$]\\ \hline
MATLAS-42 & $19.44\pm0.43$ & $0.45 \pm 0.00$ & $0.60 \pm 0.00$ & $-43.05 \pm 0.30$ & $3.11$ & $25.94$ & $25.64$ & NGC0502 & 35.9 & B,B\\
MATLAS-48 & $10.07\pm0.72$ & $0.95 \pm 0.02$ & $0.71 \pm 0.01$ & $-85.12 \pm 1.20$ & $1.72$ & $25.98$ & $24.94$ & NGC0502 & 35.9 & --,B\\
MATLAS-141 & $9.87\pm1.00$ & $0.52 \pm 0.02$ & $0.52 \pm 0.01$ & $-69.09 \pm 1.30$ & $1.73$ & $27.14$ & $26.76$ & NGC0770 & 36.7 & A,A\\
MATLAS-149 & $11.41\pm1.12$ & $0.55 \pm 0.03$ & $0.45 \pm 0.01$ & $-84.63 \pm 1.00$ & $2.00$ & $27.46$ & $27.04$ & NGC0770 & 36.7 & A,A\\
MATLAS-177 & $15.74\pm1.33$ & $0.73 \pm 0.02$ & $0.81 \pm 0.01$ & $-45.93 \pm 2.50$ & $1.69$ & $27.20$ & $26.51$ & NGC0936 & 22.4 & A,A\\
MATLAS-262 & $9.37\pm1.07$ & $1.00 \pm 0.02$ & $0.31 \pm 0.00$ & $85.97 \pm 0.51$ & $1.36$ & $26.92$ & $25.79$ & NGC1248 & 30.4 & A,A\\
MATLAS-342 & $9.65\pm0.49$ & $1.15 \pm 0.01$ & $0.76 \pm 0.00$ & $27.79 \pm 0.08$ & $1.48$ & $25.54$ & $24.16$ & NGC2481 & 32.0 & B,B\\
MATLAS-365 & $5.13\pm0.56$ & $1.00 \pm 0.02$ & $0.74 \pm 0.01$ & $-62.48 \pm 2.60$ & $0.75$ & $25.34$ & $24.22$ & NGC2577 & 30.8 & A,A\\
MATLAS-368 & $17.88\pm1.59$ & $1.15 \pm 0.02$ & $0.78 \pm 0.01$ & $-48.24 \pm 1.20$ & $2.63$ & $26.35$ & $24.98$ & NGC2577 & 30.8 & A,A\\
MATLAS-405 & $17.21\pm1.05$ & $0.78 \pm 0.01$ & $0.53 \pm 0.00$ & $50.77 \pm 0.62$ & $2.31$ & $27.08$ & $26.32$ & UGC04551 & 28.0 & A,A\\
MATLAS-478 & $19.26\pm0.52$ & $1.06 \pm 0.01$ & $0.49 \pm 0.00$ & $-65.97 \pm 0.10$ & $2.02$ & $26.15$ & $24.92$ & NGC2768 & 21.8 & A,A\\
MATLAS-585 & $11.61\pm0.45$ & $0.59 \pm 0.01$ & $0.61 \pm 0.00$ &$ 3.21 \pm 1.00 $& $1.51$ & $26.14$ & $25.66$ & IC0560 & 27.2 & B,B\\
MATLAS-658 & $12.56\pm4.57$ & $1.54 \pm 0.08$ & $0.39 \pm 0.01$ & $16.92 \pm 0.56$ & $1.99$ & $28.04$ & $25.96$ & NGC3193 & 33.1 & A,A\\
MATLAS-799 & $16.98\pm0.53$ & $0.73 \pm 0.01$ & $0.87 \pm 0.00$ & $-4.23 \pm 0.13$ & $2.00$ & $25.56$ & $24.87$ & NGC3414 & 24.5 & B,B\\
MATLAS-898 & $13.97\pm0.85$ & $1.00 \pm 0.02$ & $0.67 \pm 0.01$ & $34.34 \pm 0.94$ & $1.33$ & $25.89$ & $24.77$ & NGC3599 & 19.8 & B,B\\
MATLAS-951 & $20.83\pm1.15$ & $1.00 \pm 0.03$ & $0.45 \pm 0.00$ & $-70.71 \pm 0.20$ & $2.62$ & $26.11$ & $24.98$ & NGC3640 & 26.3 & --,B\\
MATLAS-984 & $9.45\pm0.53$ & $0.64 \pm 0.02$ & $0.73 \pm 0.01$ & $40.64 \pm 0.66$ & $1.49$ & $26.49$ & $25.93$ & NGC3665 & 33.1 & B,A\\
MATLAS-1059 & $18.94\pm1.39$ & $1.64 \pm 0.02$ & $0.60 \pm 0.00$ & $-69.50 \pm 0.58$ & $3.02$ & $26.18$ & $23.92$ & NGC3674 & 33.4 & B,B\\
MATLAS-1174 & $14.09\pm2.39$ & $1.00 \pm 0.02$ & $0.29 \pm 0.01$ & $57.06 \pm 0.57$ & $2.56$ & $27.24$ & $26.11$ & NGC4078 & 38.1 & A,A\\
MATLAS-1177 & $12.28\pm0.07$ & $1.00 \pm 0.02$ & $0.94 \pm 0.00$ & $39.37 \pm 0.01$ & $1.45$ & $25.15$ & $24.02$ & NGC4036 & 24.6 & --,B\\
MATLAS-1205 & $10.14\pm1.17$ & $0.83 \pm 0.02$ & $0.71 \pm 0.01$ & $-38.53 \pm 1.90$ & $1.89$ & $26.53$ & $25.69$ & NGC4191 & 39.2 & --,A\\
MATLAS-1216 & $8.72\pm2.75$ & $0.80 \pm 0.04$ & $1.00 \pm 0.03$ &$ 0.11 \pm 0.98$ & $1.63$ & $26.76$ & $25.97$ & NGC4191 & 39.2 & B,B\\
MATLAS-1225 & $15.01\pm1.63$ & $1.00 \pm 0.02$ & $0.54 \pm 0.01$ & $-67.13 \pm 0.80$ & $1.38$ & $26.41$ & $25.29$ & NGC4251 & 19.1 & A,A\\
MATLAS-1245 & $14.15\pm0.91$ & $0.75 \pm 0.02$ & $0.49 \pm 0.00$ & $-88.12 \pm 0.60$ & $2.13$ & $27.35$ & $26.62$ & NGC4215 & 31.5 & --,A\\
MATLAS-1246 & $15.23\pm0.88$ & $1.08 \pm 0.01$ & $0.96 \pm 0.00$ & $32.36 \pm 0.82$ & $2.70$ & $25.81$ & $24.55$ & NGC4259 & 37.2 & --,B\\
MATLAS-1248 & $15.24\pm0.22$ & $0.96 \pm 0.00$ & $0.68 \pm 0.00$ & $-64.26 \pm 0.20$ & $2.70$ & $25.42$ & $24.35$ & NGC4259 & 37.2 & --,B\\
MATLAS-1249 & $11.39\pm0.71$ & $0.86 \pm 0.02$ & $0.50 \pm 0.00$ & $59.25 \pm 0.55$ & $2.02$ & $27.08$ & $26.18$ & NGC4259 & 37.2 & --,B\\
MATLAS-1262 & $11.65\pm0.36$ & $0.88 \pm 0.01$ & $0.80 \pm 0.00$ & $-81.22 \pm 0.90$ & $1.75$ & $26.06$ & $25.13$ & NGC4215 & 31.5 & B,B\\
MATLAS-1274 & $14.41\pm0.31$ & $0.80 \pm 0.00$ & $0.79 \pm 0.00$ & $41.54 \pm 0.61$ & $2.56$ & $25.33$ & $24.53$ & NGC4259 & 37.2 & --,B\\
MATLAS-1302 & $9.85\pm0.60$ & $0.23 \pm 0.01$ & $0.28 \pm 0.00$ & $19.20 \pm 0.41$ & $1.75$ & $27.71$ & $27.64$ & NGC4259 & 37.2 & A,A\\
MATLAS-1337 & $12.56\pm0.60$ & $0.81 \pm 0.01$ & $0.61 \pm 0.00$ & $-69.72 \pm 0.60$ & $2.15$ & $25.99$ & $25.17$ & NGC4251 & 19.1 & --,A\\
MATLAS-1413 & $18.19\pm4.11$ & $1.00 \pm 0.02$ & $1.00 \pm 0.02$ & $89.23 \pm 0.43$ & $3.52$ & $26.56$ & $25.43$ & PGC042549 & 40.7 & B,B\\
MATLAS-1493 & $9.69\pm1.01$ & $0.99 \pm 0.03$ & $0.62 \pm 0.01$ & $73.25 \pm 0.38$ & $1.86$ & $26.67$ & $25.55$ & NGC4690 & 40.2 & --,A\\
MATLAS-1494 & $12.49\pm0.66$ & $0.65 \pm 0.01$ & $0.64 \pm 0.00$ & $-80.86 \pm 0.50$ & $2.39$ & $26.46$ & $25.89$ & NGC4690 & 40.2 & --,A\\
MATLAS-1534 & $5.22\pm0.63$ & $1.00 \pm 0.02$ & $0.61 \pm 0.01$ & $-25.30 \pm 1.90$ & $0.98$ & $26.06$ & $24.94$ & NGC5198 & 39.6 & A,A\\
MATLAS-1550 & $10.58\pm0.75$ & $0.86 \pm 0.02$ & $0.51 \pm 0.00$ & $-18.97 \pm 0.60$ & $1.59$ & $26.85$ & $25.95$ & NGC5308 & 31.5 & A,A\\
MATLAS-1558 & $14.16\pm0.99$ & $0.68 \pm 0.02$ & $0.62 \pm 0.01$ & $-33.43 \pm 1.10$ & $2.13$ & $27.36$ & $26.74$ & NGC5308 & 31.5 & A,A\\
MATLAS-1589 & $11.74\pm1.52$ & $1.00 \pm 0.02$ & $0.53 \pm 0.01$ & $-27.34 \pm 1.30$ & $1.70$ & $26.74$ & $25.61$ & NGC5322 & 30.3 & A,A\\
MATLAS-1606 & $9.77\pm0.56$ & $1.00 \pm 0.02$ & $0.85 \pm 0.01$ & $58.20 \pm 0.56$ & $1.73$ & $25.31$ & $24.19$ & NGC5355 & 37.1 & --,B\\
MATLAS-1615 & $16.11\pm0.64$ & $0.79 \pm 0.01$ & $0.77 \pm 0.00$ & $-49.48 \pm 1.00$ & $2.85$ & $26.30$ & $25.52$ & NGC5355 & 37.1 & --,A\\
MATLAS-1616 & $17.09\pm1.33$ & $0.98 \pm 0.02$ & $0.58 \pm 0.00$ & $-47.39 \pm 0.70$ & $2.45$ & $26.82$ & $25.73$ & NGC5379 & 30.0 & B,B\\
MATLAS-1630 & $13.16\pm1.40$ & $1.19 \pm 0.03$ & $0.56 \pm 0.01$ &$ 0.27 \pm 0.43$ & $1.91$ & $26.89$ & $25.44$ & NGC5322 & 30.3 & B,B\\
MATLAS-1647 & $7.52\pm0.64$ & $0.54 \pm 0.03$ & $0.33 \pm 0.01$ & $86.86 \pm 0.64$ & $1.33$ & $27.19$ & $26.78$ & NGC5355 & 37.1 & A,A\\
MATLAS-1779 & $14.50\pm1.34$ & $1.00 \pm 0.02$ & $0.47 \pm 0.01$ & $22.12 \pm 0.12$ & $2.68$ & $26.50$ & $25.37$ & NGC5493 & 38.8 & A,A\\
MATLAS-1794 & $9.02\pm0.21$ & $1.05 \pm 0.01$ & $0.32 \pm 0.00$ & $44.44 \pm 0.10$ & $1.23$ & $26.06$ & $24.84$ & NGC5507 & 28.5 & A,A\\
MATLAS-1824 & $13.31\pm0.63$ & $0.98 \pm 0.01$ & $0.67 \pm 0.00$ & $71.43 \pm 0.74$ & $2.93$ & $25.99$ & $24.89$ & NGC5557 & 38.8 & --,A\\
MATLAS-1847 & $15.58\pm0.84$ & $0.32 \pm 0.01$ & $0.52 \pm 0.01$ & $20.84 \pm 0.76$ & $1.73$ & $27.27$ & $27.13$ & NGC5574 & 23.2 & --,A\\
MATLAS-1855 & $11.72\pm0.74$ & $0.80 \pm 0.01$ & $0.73 \pm 0.01$ & $-44.86 \pm 1.40$ & $1.30$ & $26.35$ & $25.55$ & NGC5574 & 23.2 & --,A\\
MATLAS-1865 & $11.17\pm0.28$ & $0.71 \pm 0.01$ & $0.35 \pm 0.00$ & $-71.96 \pm 0.10$ & $1.44$ & $25.98$ & $25.32$ & NGC5631 & 27.0 & A,A\\
MATLAS-1907 & $12.83\pm0.28$ & $0.90 \pm 0.01$ & $0.45 \pm 0.00$ & $-48.91 \pm 0.10$ & $1.49$ & $25.77$ & $24.82$ & IC1024 & 24.2 & B,B\\
MATLAS-1957 & $7.64\pm0.33$ & $1.00 \pm 0.02$ & $0.80 \pm 0.00$ & $-74.43 \pm 1.30$ & $1.14$ & $24.64$ & $23.51$ & NGC5813 & 31.3 & --,B\\
MATLAS-1975 & $13.00\pm1.80$ & $1.58 \pm 0.04$ & $0.80 \pm 0.01$ & $-40.81 \pm 2.10$ & $1.64$ & $26.71$ & $24.54$ & NGC5831 & 26.4 & A,A\\
MATLAS-1985 & $15.36\pm1.30$ & $1.00 \pm 0.02$ & $0.72 \pm 0.01$ & $38.87 \pm 0.12$ & $1.94$ & $26.04$ & $24.92$ & NGC5831 & 26.4 & B,B\\
MATLAS-1991 & $12.63\pm0.26$ & $0.73 \pm 0.01$ & $0.68 \pm 0.00$ & $10.37 \pm 0.41$ & $1.53$ & $25.48$ & $24.79$ & NGC5845 & 25.2 &  --,B\\
MATLAS-1996 & $14.87\pm0.36$ & $1.00 \pm 0.01$ & $0.29 \pm 0.01$ & $-62.51 \pm 0.80$ & $1.56$ & $27.82$ & $26.69$ & NGC5838 & 21.8 & --,A\\
MATLAS-2019 & $15.74\pm0.36$ & $0.61 \pm 0.06$ & $0.97 \pm 0.03$ & $-78.49 \pm 5.30$ & $1.51$ & $25.85$ & $25.34$ & NGC5845 & 25.2 & B,B\\
MATLAS-2021 & $17.79\pm0.45$ & $0.76 \pm 0.03$ & $0.84 \pm 0.02$ & $-40.85 \pm 6.60$ & $1.86$ & $27.67$ & $26.94$ & NGC5838 & 21.8 & A,A\\
MATLAS-2103 & $15.34\pm0.21$ & $0.81 \pm 0.00$ & $0.69 \pm 0.00$ & $-89.43 \pm 0.20$ & $2.62$ & $25.90$ & $25.10$ & NGC6014 & 35.8 & --,B\\
MATLAS-2184 & $9.46\pm1.04$ & $1.00 \pm 0.03$ & $0.74 \pm 0.01$ & $83.62 \pm 0.52$ & $1.33$ & $26.55$ & $25.42$ & NGC7465 & 29.3 & A,A\\
\hline
\end{tabular}
\begin{tablenotes}
      \small
      \item \textbf{Note.} Columns are: (1) Galaxy ID; (2) \texttt{GALFITM} effective radius in arcsec; (3) \texttt{GALFITM} S\'ersic index; (4) \texttt{GALFITM} axis ratio; (5) \texttt{GALFITM} position angle; (6) Effective radius in kpc; (7) Mean surface brightness; (8) Central surface brightness; (9) Candidate host of UDGs; (10) Distance to host (and assumed distance to UDG); (11) \texttt{KMeans} clustering algorithm class (first class is for the determination using $N_{\rm GC}$ -- only 38 galaxies; the second is for the classification without $N_{\rm GC}$ -- all 59 UDGs. 
\end{tablenotes}
\label{tab:morphology}
\end{threeparttable}}
\end{table*}

\subsection{Stellar population properties from \texttt{Prospector}}
\label{sec:appendix_prospector}

\renewcommand{\arraystretch}{1.3}
\begin{table*}
\scalebox{0.78}{
\begin{threeparttable}
\caption{\texttt{Prospector} stellar population properties of the MATLAS UDGs.}
\begin{tabular}{ccccccccccccccccccc} \hline
\multirow{2}{*}{ID} & \multirow{2}{*}{log($M_{\star}/M_{\odot}$)} & [M/H] & $\tau$ & $t_M$ & $A_v$ & $M_{\star}/L_{V}$ \\ 
  &  & [dex] & [Gyr] & [Gyr] & [mag] & [$M_{\odot}/L_{\odot,V}$]  \\ \hline
MATLAS-42 & $8.17^{+0.08}_{-0.10}$ & $-1.34^{+0.30}_{-0.19}$ & $1.57^{+1.70}_{-1.08}$ & $5.97^{+5.46}_{-3.59}$ & $0.07^{+0.09}_{-0.05}$ & 0.97 \\
MATLAS-48 & $7.89^{+0.07}_{-0.10}$ & $-1.31^{+0.30}_{-0.20}$ & $1.19^{+1.13}_{-0.76}$ & $8.90^{+3.50}_{-4.30}$ & $0.07^{+0.08}_{-0.05}$ & 1.99 \\
MATLAS-141 & $7.31^{+0.14}_{-0.16}$ & $-1.06^{+0.58}_{-0.40}$ & $2.85^{+3.79}_{-2.01}$ & $5.13^{+5.78}_{-2.94}$ & $0.15^{+0.20}_{-0.11}$ & 1.44 \\
MATLAS-149 & $7.25^{+0.12}_{-0.14}$ & $-1.42^{+0.26}_{-0.13}$ & $3.32^{+3.36}_{-2.33}$ & $4.04^{+5.87}_{-2.16}$ & $0.04^{+0.06}_{-0.03}$ & 1.25 \\
MATLAS-177 & $7.49^{+0.09}_{-0.13}$ & $-1.19^{+0.41}_{-0.28}$ & $1.43^{+1.63}_{-0.93}$ & $8.34^{+3.84}_{-4.33}$ & $0.16^{+0.14}_{-0.10}$ & 2.52 \\
MATLAS-262 & $7.40^{+0.19}_{-0.19}$ & $-1.21^{+0.68}_{-0.54}$ & $2.74^{+3.53}_{-1.98}$ & $6.02^{+5.46}_{-3.31}$ & $0.50^{+0.30}_{-0.30}$ & 2.31 \\
MATLAS-342 & $7.81^{+0.09}_{-0.10}$ & $-1.22^{+0.33}_{-0.26}$ & $1.48^{+1.65}_{-0.98}$ & $7.49^{+4.51}_{-4.20}$ & $0.12^{+0.11}_{-0.08}$ & 1.77 \\
MATLAS-365 & $7.12^{+0.13}_{-0.16}$ & $-1.16^{+0.51}_{-0.32}$ & $2.86^{+3.41}_{-1.98}$ & $5.33^{+5.69}_{-2.93}$ & $0.15^{+0.19}_{-0.11}$ & 0.96 \\
MATLAS-368 & $7.23^{+0.08}_{-0.10}$ & $-1.42^{+0.22}_{-0.13}$ & $2.59^{+4.35}_{-2.29}$ & $4.44^{+0.89}_{-0.16}$ & $0.03^{+0.01}_{-0.00}$ & 0.26 \\
MATLAS-405 & $7.64^{+0.09}_{-0.12}$ & $-1.21^{+0.41}_{-0.27}$ & $1.57^{+1.77}_{-1.06}$ & $7.90^{+4.24}_{-4.19}$ & $0.12^{+0.13}_{-0.08}$ & 1.75 \\
MATLAS-478 & $8.05^{+0.04}_{-0.06}$ & $-0.69^{+0.39}_{-0.28}$ & $0.77^{+0.56}_{-0.48}$ & $10.92^{+1.99}_{-2.98}$ & $0.22^{+0.10}_{-0.13}$ & 2.43 \\
MATLAS-585 & $7.53^{+0.07}_{-0.08}$ & $-1.36^{+0.27}_{-0.16}$ & $1.12^{+1.18}_{-0.69}$ & $8.91^{+3.44}_{-4.31}$ & $0.06^{+0.07}_{-0.04}$ & 1.24 \\
MATLAS-658 & $7.39^{+0.26}_{-0.24}$ & $-1.08^{+0.71}_{-0.62}$ & $3.18^{+3.76}_{-2.32}$ & $5.28^{+5.85}_{-3.03}$ & $1.17^{+0.31}_{-0.33}$ & 3.16 \\
MATLAS-799 & $7.87^{+0.05}_{-0.06}$ & $-1.56^{+0.06}_{-0.03}$ & $0.69^{+0.84}_{-0.43}$ & $3.67^{+2.32}_{-1.76}$ & $0.01^{+0.01}_{-0.00}$ & 0.97 \\
MATLAS-898 & $7.35^{+0.09}_{-0.10}$ & $-1.51^{+0.13}_{-0.07}$ & $1.83^{+2.92}_{-1.39}$ & $3.85^{+6.63}_{-2.31}$ & $0.01^{+0.02}_{-0.01}$ & 0.90 \\
MATLAS-951 & $8.25^{+0.10}_{-0.15}$ & $-1.33^{+0.56}_{-0.45}$ & $0.88^{+0.96}_{-0.57}$ & $9.19^{+3.32}_{-4.54}$ & $0.37^{+0.17}_{-0.22}$ & 3.46 \\
MATLAS-984 & $7.34^{+0.09}_{-0.09}$ & $-1.50^{+0.15}_{-0.08}$ & $1.50^{+2.01}_{-1.04}$ & $5.70^{+5.58}_{-3.21}$ & $0.02^{+0.03}_{-0.01}$ & 1.20 \\
MATLAS-1059 & $7.84^{+0.10}_{-0.12}$ & $-1.50^{+0.15}_{-0.07}$ & $2.07^{+2.84}_{-1.47}$ & $4.47^{+6.32}_{-2.60}$ & $0.02^{+0.03}_{-0.01}$ & 0.77 \\
MATLAS-1174 & $7.57^{+0.15}_{-0.18}$ & $-1.23^{+0.44}_{-0.27}$ & $2.39^{+3.25}_{-1.65}$ & $6.12^{+5.31}_{-3.48}$ & $0.09^{+0.13}_{-0.07}$ & 1.38 \\
MATLAS-1177 & $6.92^{+0.18}_{-0.22}$ & $-0.92^{+0.61}_{-0.71}$ & $6.33^{+2.50}_{-2.95}$ & $2.70^{+3.20}_{-1.43}$ & $0.37^{+0.12}_{-0.12}$ & 0.25 \\
MATLAS-1205 & $7.37^{+0.13}_{-0.14}$ & $-1.39^{+0.30}_{-0.16}$ & $3.11^{+3.52}_{-2.25}$ & $5.27^{+5.81}_{-2.76}$ & $0.04^{+0.07}_{-0.03}$ & 0.83 \\
MATLAS-1216 & $7.45^{+0.13}_{-0.21}$ & $-0.94^{+0.60}_{-0.69}$ & $0.96^{+1.41}_{-0.63}$ & $9.00^{+3.50}_{-5.24}$ & $0.04^{+0.07}_{-0.03}$ & 0.66 \\
MATLAS-1225 & $7.30^{+0.12}_{-0.16}$ & $-1.35^{+0.32}_{-0.18}$ & $1.48^{+2.40}_{-1.01}$ & $7.28^{+4.75}_{-4.39}$ & $0.07^{+0.09}_{-0.05}$ & 1.21 \\
MATLAS-1245 & $7.39^{+0.09}_{-0.12}$ & $-1.13^{+0.46}_{-0.33}$ & $1.71^{+2.43}_{-1.16}$ & $7.66^{+4.35}_{-4.37}$ & $0.18^{+0.22}_{-0.13}$ & 1.46 \\
MATLAS-1246 & $8.17^{+0.08}_{-0.08}$ & $-1.49^{+0.16}_{-0.08}$ & $1.04^{+0.95}_{-0.63}$ & $7.92^{+4.01}_{-3.74}$ & $0.03^{+0.04}_{-0.02}$ & 1.31 \\
MATLAS-1248 & $8.49^{+0.02}_{-0.05}$ & $-1.28^{+0.22}_{-0.11}$ & $0.50^{+0.53}_{-0.30}$ & $11.81^{+1.31}_{-2.86}$ & $0.07^{+0.05}_{-0.05}$ & 1.93 \\
MATLAS-1249 & $6.61^{+0.24}_{-0.27}$ & $-1.05^{+0.86}_{-0.71}$ & $6.30^{+2.55}_{-2.99}$ & $1.15^{+1.80}_{-0.64}$ & $1.06^{+0.21}_{-0.20}$ & 0.12 \\
MATLAS-1262 & $7.89^{+0.06}_{-0.08}$ & $-1.27^{+0.34}_{-0.23}$ & $0.92^{+0.99}_{-0.61}$ & $9.40^{+3.13}_{-4.34}$ & $0.14^{+0.10}_{-0.09}$ & 2.08 \\
MATLAS-1274 & $8.26^{+0.08}_{-0.07}$ & $-1.51^{+0.13}_{-0.07}$ & $0.96^{+1.07}_{-0.59}$ & $5.97^{+4.51}_{-2.98}$ & $0.02^{+0.03}_{-0.01}$ & 1.17 \\
MATLAS-1302 & $7.08^{+0.11}_{-0.15}$ & $-1.01^{+0.50}_{-0.41}$ & $1.70^{+2.67}_{-1.18}$ & $7.66^{+4.50}_{-4.28}$ & $0.22^{+0.19}_{-0.15}$ & 1.48 \\
MATLAS-1337 & $7.04^{+0.15}_{-0.17}$ & $-0.91^{+0.70}_{-0.47}$ & $4.98^{+3.41}_{-2.81}$ & $2.43^{+3.38}_{-1.32}$ & $0.12^{+0.13}_{-0.09}$ & 0.64 \\
MATLAS-1413 & $7.85^{+0.21}_{-0.27}$ & $-1.40^{+0.60}_{-0.20}$ & $4.38^{+3.60}_{-3.00}$ & $4.08^{+5.69}_{-2.37}$ & $0.38^{+0.35}_{-0.26}$ & 0.88 \\
MATLAS-1493 & $7.21^{+0.15}_{-0.22}$ & $-1.03^{+0.70}_{-0.40}$ & $4.40^{+3.84}_{-2.89}$ & $3.57^{+5.28}_{-2.25}$ & $0.31^{+0.31}_{-0.21}$ & 0.66 \\
MATLAS-1494 & $8.00^{+0.11}_{-0.17}$ & $-1.10^{+0.56}_{-0.56}$ & $0.90^{+1.06}_{-0.56}$ & $9.15^{+3.29}_{-4.71}$ & $0.04^{+0.06}_{-0.03}$ & 1.80 \\
MATLAS-1534 & $7.65^{+0.20}_{-0.33}$ & $-0.76^{+0.52}_{-0.77}$ & $1.00^{+1.50}_{-0.65}$ & $9.00^{+3.42}_{-5.11}$ & $0.31^{+0.33}_{-0.22}$ & 3.14 \\
MATLAS-1550 & $7.47^{+0.10}_{-0.14}$ & $-1.16^{+0.43}_{-0.30}$ & $1.37^{+1.87}_{-0.93}$ & $8.20^{+4.08}_{-4.53}$ & $0.18^{+0.18}_{-0.12}$ & 1.98 \\
MATLAS-1558 & $7.33^{+0.14}_{-0.14}$ & $-1.29^{+0.40}_{-0.23}$ & $2.22^{+2.55}_{-1.56}$ & $5.97^{+5.53}_{-3.41}$ & $0.07^{+0.10}_{-0.05}$ & 1.27 \\
MATLAS-1589 & $7.11^{+0.17}_{-0.22}$ & $-0.94^{+0.60}_{-0.71}$ & $1.54^{+2.00}_{-1.01}$ & $7.82^{+4.28}_{-4.50}$ & $0.21^{+0.22}_{-0.14}$ & 1.20 \\
MATLAS-1606 & $8.09^{+0.09}_{-0.11}$ & $-1.35^{+0.30}_{-0.18}$ & $1.17^{+1.32}_{-0.75}$ & $8.63^{+3.67}_{-4.60}$ & $0.08^{+0.09}_{-0.06}$ & 1.70 \\
MATLAS-1615 & $7.87^{+0.11}_{-0.14}$ & $-1.09^{+0.54}_{-0.36}$ & $2.95^{+2.74}_{-2.01}$ & $4.63^{+6.09}_{-2.66}$ & $0.15^{+0.16}_{-0.11}$ & 0.94 \\
MATLAS-1616 & $7.73^{+0.11}_{-0.12}$ & $-1.26^{+0.40}_{-0.24}$ & $1.54^{+2.06}_{-1.03}$ & $7.77^{+4.33}_{-4.43}$ & $0.13^{+0.14}_{-0.09}$ & 1.51 \\
MATLAS-1630 & $7.28^{+0.08}_{-0.11}$ & $-1.34^{+0.30}_{-0.18}$ & $1.12^{+1.22}_{-0.72}$ & $9.07^{+3.35}_{-4.34}$ & $0.06^{+0.08}_{-0.05}$ & 0.94 \\
MATLAS-1647 & $6.96^{+0.11}_{-0.14}$ & $-1.04^{+0.55}_{-0.40}$ & $1.93^{+2.80}_{-1.33}$ & $7.09^{+4.79}_{-3.97}$ & $0.33^{+0.23}_{-0.21}$ & 1.22 \\
MATLAS-1779 & $7.91^{+0.14}_{-0.17}$ & $-0.91^{+0.57}_{-0.47}$ & $2.52^{+3.26}_{-1.79}$ & $5.63^{+5.79}_{-3.35}$ & $0.36^{+0.32}_{-0.23}$ & 1.37 \\
MATLAS-1794 & $7.40^{+0.12}_{-0.20}$ & $-1.24^{+0.60}_{-0.49}$ & $0.98^{+1.69}_{-0.63}$ & $8.83^{+3.56}_{-5.20}$ & $0.01^{+0.02}_{-0.01}$ & 1.09 \\
MATLAS-1824 & $7.95^{+0.12}_{-0.13}$ & $-0.88^{+0.39}_{-0.22}$ & $1.70^{+2.32}_{-1.17}$ & $6.79^{+5.02}_{-4.07}$ & $0.07^{+0.09}_{-0.05}$ & 1.14 \\
MATLAS-1847 & $7.36^{+0.10}_{-0.12}$ & $-1.08^{+0.43}_{-0.36}$ & $1.67^{+2.57}_{-1.16}$ & $7.38^{+4.65}_{-4.26}$ & $0.20^{+0.19}_{-0.14}$ & 1.91 \\
MATLAS-1855 & $7.28^{+0.12}_{-0.15}$ & $-1.20^{+0.50}_{-0.30}$ & $2.45^{+2.92}_{-1.71}$ & $5.47^{+5.76}_{-3.33}$ & $0.16^{+0.21}_{-0.11}$ & 1.19 \\
MATLAS-1865 & $7.43^{+0.09}_{-0.11}$ & $-1.48^{+0.18}_{-0.09}$ & $2.49^{+2.83}_{-1.92}$ & $3.80^{+6.71}_{-2.37}$ & $0.02^{+0.03}_{-0.01}$ & 1.21 \\
MATLAS-1907 & $7.55^{+0.18}_{-0.22}$ & $-1.68^{+0.37}_{-0.23}$ & $1.31^{+4.41}_{-0.91}$ & $7.34^{+4.68}_{-4.85}$ & $0.00^{+0.00}_{-0.00}$ & 0.92 \\
MATLAS-1957 & $8.00^{+0.05}_{-0.07}$ & $-1.47^{+0.12}_{-0.09}$ & $0.84^{+0.74}_{-0.52}$ & $10.04^{+2.64}_{-3.90}$ & $0.02^{+0.03}_{-0.01}$ & 1.65 \\
MATLAS-1975 & $7.58^{+0.13}_{-0.15}$ & $-0.76^{+0.68}_{-0.57}$ & $2.74^{+3.25}_{-1.94}$ & $6.09^{+5.36}_{-3.27}$ & $0.37^{+0.30}_{-0.24}$ & 2.04 \\
MATLAS-1985 & $7.96^{+0.10}_{-0.13}$ & $-1.40^{+0.25}_{-0.14}$ & $1.25^{+1.65}_{-0.82}$ & $7.94^{+4.10}_{-4.42}$ & $0.05^{+0.08}_{-0.04}$ & 1.92 \\
MATLAS-1991 & $7.88^{+0.04}_{-0.06}$ & $-1.35^{+0.19}_{-0.16}$ & $0.75^{+0.70}_{-0.48}$ & $10.84^{+2.13}_{-3.75}$ & $0.09^{+0.06}_{-0.06}$ & 1.53 \\
MATLAS-1996 & $6.69^{+0.18}_{-0.23}$ & $-0.79^{+0.74}_{-0.58}$ & $4.99^{+3.14}_{-2.99}$ & $3.50^{+4.71}_{-2.13}$ & $0.37^{+0.40}_{-0.26}$ & 0.83 \\
MATLAS-2019 & $8.01^{+0.03}_{-0.05}$ & $-1.40^{+0.10}_{-0.12}$ & $0.67^{+0.63}_{-0.42}$ & $11.22^{+1.79}_{-3.23}$ & $0.04^{+0.04}_{-0.03}$ & 1.88 \\
MATLAS-2021 & $7.08^{+0.16}_{-0.20}$ & $-0.88^{+0.65}_{-0.52}$ & $3.95^{+3.54}_{-2.74}$ & $4.55^{+5.69}_{-2.81}$ & $0.41^{+0.39}_{-0.28}$ & 1.22 \\
MATLAS-2103 & $8.21^{+0.05}_{-0.06}$ & $-1.40^{+0.14}_{-0.13}$ & $0.92^{+0.69}_{-0.57}$ & $9.79^{+2.81}_{-3.84}$ & $0.06^{+0.05}_{-0.04}$ & 1.66 \\
MATLAS-2184 & $7.48^{+0.18}_{-0.20}$ & $-0.77^{+0.79}_{-0.60}$ & $3.11^{+3.71}_{-2.11}$ & $4.64^{+5.98}_{-2.83}$ & $0.38^{+0.46}_{-0.27}$ & 2.09 \\
  \hline
\end{tabular}
\begin{tablenotes}
      \small
      \item \textbf{Note.} Columns are: (1) Galaxy ID; (2) \texttt{Prospector} stellar mass; (3) \texttt{Prospector} metallicity; (4) \texttt{Prospector} Star formation timescale; (5)  \texttt{Prospector} Mass-weighted Age; (6) \texttt{Prospector} Dust attenuation; (7) Mass-to-light ratio. 
\end{tablenotes}
\label{tab:stellarpops}
\end{threeparttable}}
\end{table*}

\section{The effects of underestimating uncertainties}
\label{sec:appendix_galfit_unders}

Galaxy fitting codes are known for severely underestimating the uncertainties of their recovered parameters.
In this Appendix, we show the effects on the stellar populations properties obtained with \texttt{Prospector} if the photometric uncertainties from \texttt{GALFITM} were underestimated by 10\%, 20\%, 50\% and 100\%. 

\begin{figure*}
    \centering
    \includegraphics[width=\columnwidth]{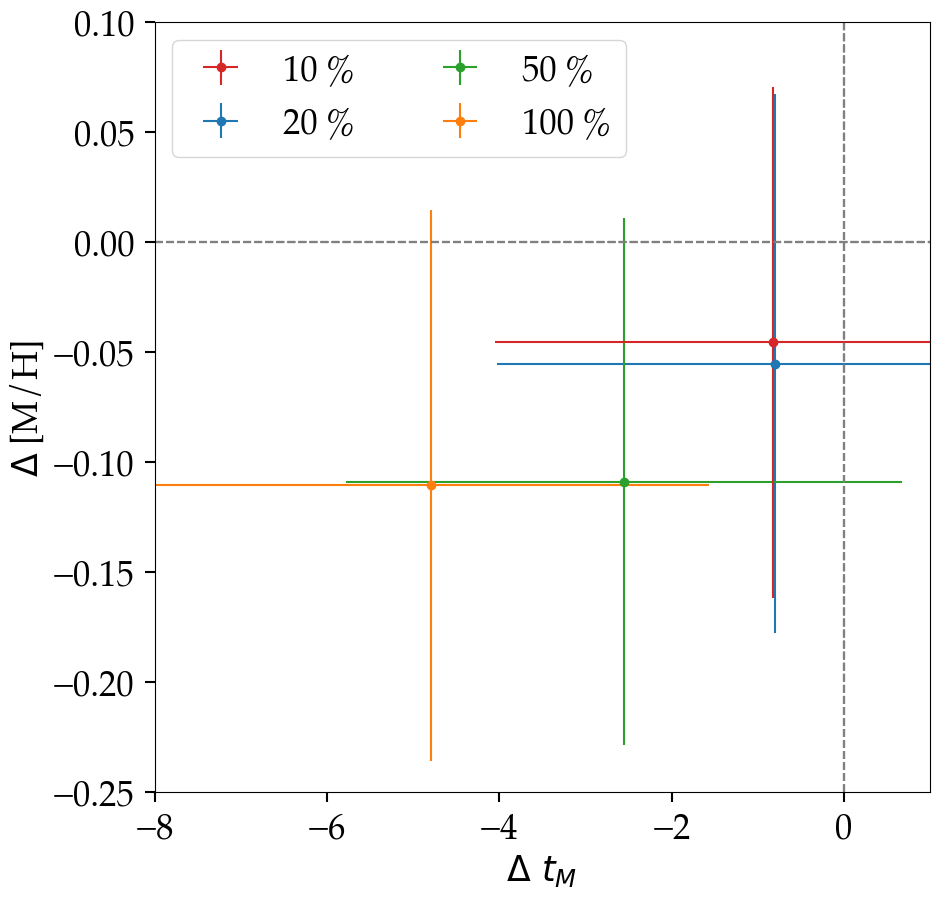}
    \includegraphics[width=\columnwidth]{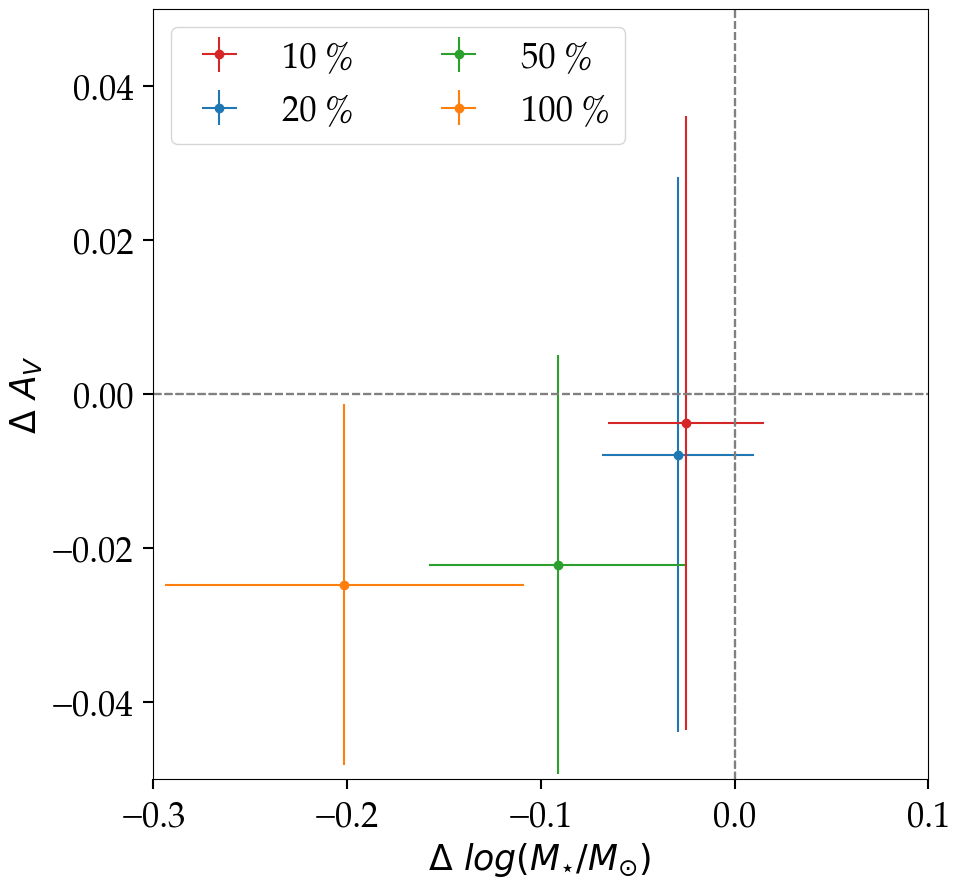}
    \caption{Differences in the recovered stellar populations of MATLAS-2019 if the photometric uncertainty in the optical obtained from \texttt{GALFITM} was underestimated by 10\% (red), 20\% (blue), 50\% (green) and 100\% (orange). \textit{Left:} Differences in metallicity and mass-weighted ages obtained. \textit{Right:} Differences in stellar mass and dust attenuation. We conclude that the photometry being underestimated by 50\% or 100\% can severely change the ages of the galaxies, reaching a change of 5 Gyr. The maximum change in metallicity is 0.11 dex, in stellar mass is 0.2 dex and in $A_V$ is 0.02. Due to the small changes in mass and metallicity, the main conclusions of the paper regarding the bimodality in the mass-metallicity plane remain even if the uncertainties are found to be extremely underestimated.}
    \label{fig:comp_underestimated}
\end{figure*}

In Fig. \ref{fig:comp_underestimated}, we compare the stellar populations properties of MATLAS-2019 using the nominal uncertainties coming from \texttt{GALFITM} and if they were underestimated. We can see that if the uncertainties were underestimated by 10 or 20\%, the recovered stellar populations barely change. However, if they were severely underestimated, such as by 50 or 100\%, the changes start to get important. We emphasise, however, that we do not expect the uncertainties to be this highly underestimated. The biggest of these changes is in the mass weighted ages reaching a change of 5 Gyr. The biggest change in metallicity is 0.1 dex, 0.2 dex in mass and 0.02 in $A_V$. 
The minor changes in metallicity and mass give us confidence that even if the uncertainties are underestimated, the major conclusions in this paper (i.e., the bimodality in the mass--metallicity plane) would mantain.

\section{Alternative view of Fig. 7}

\begin{figure*}
    \includegraphics[width=\textwidth]{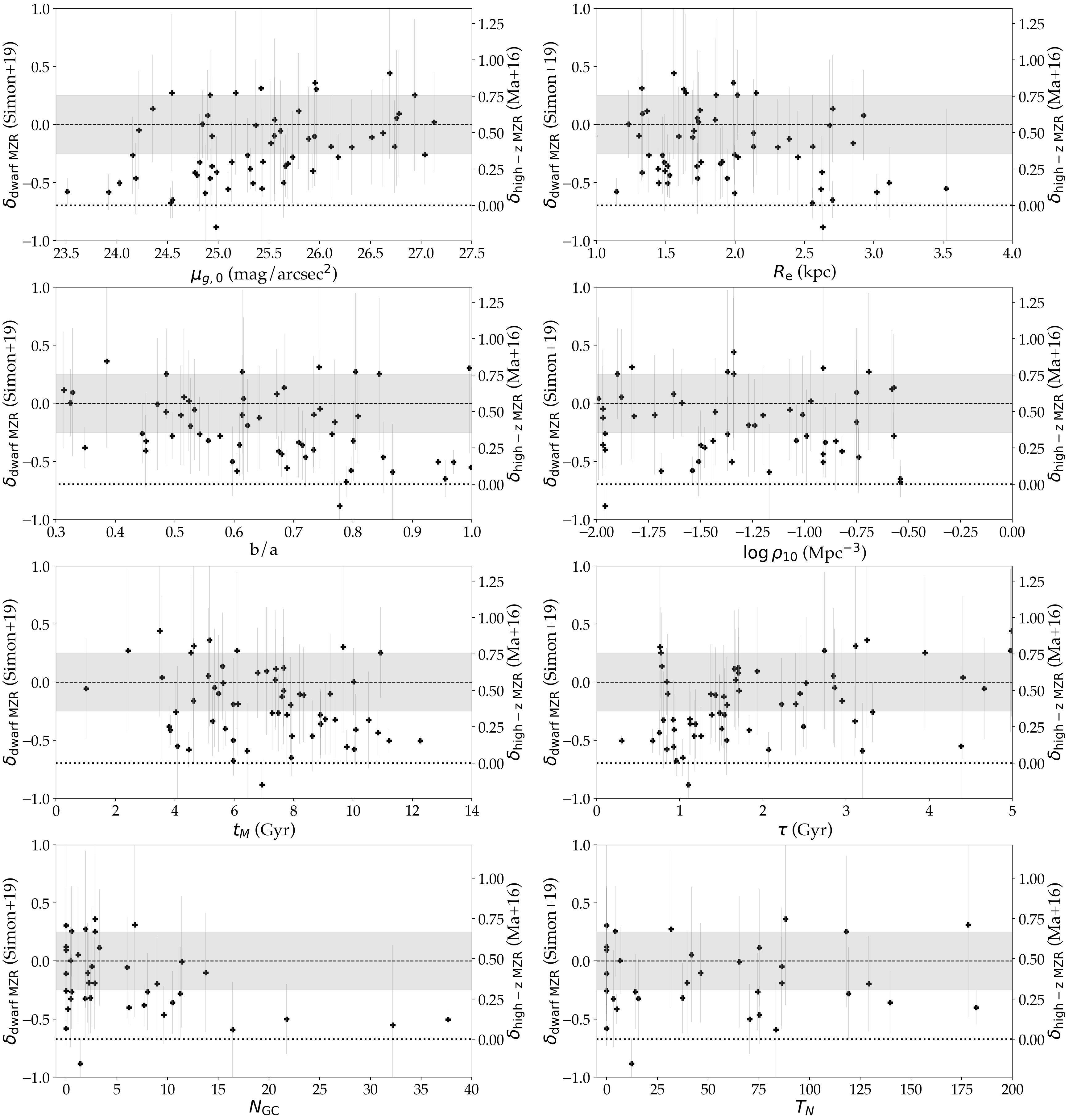}
    \caption{Alternative view of Fig. \ref{fig:MZR_matlasonly}. Distribution of the physical properties of the MATLAS UDGs in the stellar mass--metallicity plane. Plus signs show the results obtained from SED fitting with \texttt{PROSPECTOR} for the 59 UDGs in our sample. The \protect\cite{Simon_19} MZR for classical Local Group dwarf galaxies is shown with the dashed black line. The dash-dotted line is the evolving MZR at $z=2.2$ from \protect\cite{Ma_15}. In all panels, the properties are plotted against distance to the dwarf (left-hand side y-axis) and high-redshift MZRs (right-hand side y-axis). \textit{Top-row:} Surface brightness (left) and effective radii (right). \textit{Second row}: Axis ratio and local environment \protect\citep{Marleau_21}, respectively. \textit{Third row:} Mass-weighted ages and star formation timescales. \textit{Fourth row:} Number of GCs (Marleau et al. subm.) and GC specific frequencies (per unit stellar mass), respectively. Results are as described in the caption of Fig. \ref{fig:MZR_matlasonly}}
    \label{fig:direct_plots}
\end{figure*}

In Fig. \ref{fig:MZR_matlasonly} we show the MATLAS UDGs overlaid in the mass-metallcity plane in an attempt to compare their positioning to known MZRs and to start inferring some of their likely formation histories. We colour-code the MATLAS UDGs by several key parameters and show that these galaxies distill into two main types: some that are consistent with the classical dwarf MZR and some that are consistent with the high-redshift MZR. As the colour gradients may be hard to visualise in some cases, in this appendix we provide an alternative view of Fig. \ref{fig:MZR_matlasonly}, shown in Fig. \ref{fig:direct_plots}. In this case, the key properties are directly plotted against their distance to both the dwarf MZR and the high-redshift MZR.

% If you want to present additional material which would interrupt the flow of the main paper,
% it can be placed in an Appendix which appears after the list of references.

%%%%%%%%%%%%%%%%%%%%%%%%%%%%%%%%%%%%%%%%%%%%%%%%%%

% Don't change these lines
\bsp	% typesetting comment
\label{lastpage}
\end{document}